\documentclass[journal,comsoc]{IEEEtran}
\IEEEoverridecommandlockouts
\usepackage{cite}
\usepackage{amsmath,amssymb,amsfonts}
\usepackage{dsfont}
\usepackage{float}
\usepackage{graphicx,lipsum,multicol}
\usepackage{caption}
\usepackage{subcaption}
\usepackage{mwe}
\usepackage{textcomp}
\usepackage{xcolor}
\usepackage{amsmath}
\usepackage[english]{babel}
\usepackage[utf8]{inputenc}
\usepackage{mathbbol}
\usepackage{amssymb}  
\usepackage{authblk}
\usepackage{stackengine}
\usepackage{stfloats}
\usepackage{multirow}
\usepackage[linesnumbered,ruled,vlined]{algorithm2e}
\usepackage[noend]{algpseudocode}

\def\BibTeX{{\rm B\kern-.05em{\sc i\kern-.025em b}\kern-.08em
    T\kern-.1667em\lower.7ex\hbox{E}\kern-.125emX}}
\newcommand{\MYheader}{\smash{\scriptsize
\hfil\parbox[t][\height][t]{\textwidth}{\centering
This article has been accepted for publication in a future issue of this journal, but has not been fully edited. Content may change prior to final publication.\\
Citation information: DOI 10.1109/JIOT.2020.3009540, IEEE Internet of
Things Journal}\hfil\hbox{}}}
\newcommand{\MYfooter}{\smash{\scriptsize
\hfil\parbox[t][\height][t]{\textwidth}{\centering
2327-4662 (c) 2020 IEEE. Personal use is permitted, but republication/redistribution requires IEEE permission. \\
See http://www.ieee.org/publications\_standards/publications/rights/index.html for more information.}\hfil\hbox{}}}

\makeatletter

\def\ps@headings{%
\def\@oddhead{\MYheader}%
\def\@evenhead{\MYheader}%
\def\@oddfoot{\MYfooter}%
\def\@evenfoot{\MYfooter}}

\def\ps@IEEEtitlepagestyle{%
\def\@oddhead{\MYheader}%
\def\@evenhead{\MYheader}%
\def\@oddfoot{\MYfooter}%
\def\@evenfoot{\MYfooter}}

\makeatother

\begin{document}

\title{Heterogeneous Task Offloading and Resource Allocations via Deep Recurrent Reinforcement Learning in Partial Observable Multi-Fog Networks}
	
\author{Jungyeon~Baek,~\IEEEmembership{Member,~IEEE} and
        Georges~Kaddoum,~\IEEEmembership{Member,~IEEE}
\thanks{J. Baek and G. Kaddoum are with the Department of Electrical Engineering, École de Technologie Supérieure, Montréal, QC, Canada. (e-mails: jungyeon.baek.1@ens.etsmtl.ca, georges.kaddoum@etsmtl.ca)}}

\maketitle

\begin{abstract}
As wireless services and applications become more sophisticated and require faster and higher-capacity networks, there is a need for an efficient management of the execution of increasingly complex tasks based on the requirements of each application. In this regard, fog computing enables the integration of virtualized servers into networks and brings cloud services closer to end devices. In contrast to the cloud server, the computing capacity of fog nodes is limited and thus a single fog node might not be capable of computing-intensive tasks. In this context, task offloading can be particularly useful at the fog nodes by selecting the suitable nodes and proper resource management while guaranteeing the Quality-of-Service (QoS) requirements of the users. This paper studies the design of a joint task offloading and resource allocation control for heterogeneous service tasks in multi-fog nodes systems. This problem is formulated as a partially observable stochastic game, in which each fog node cooperates to maximize the aggregated local rewards while the nodes only have access to local observations. To deal with partial observability, we apply a deep recurrent Q-network (DRQN) approach to approximate the optimal value functions. The solution is then compared to a deep Q-network (DQN) and deep convolutional Q-network (DCQN) approach to evaluate the performance of different neural networks. Moreover, to guarantee the convergence and accuracy
of the neural network, an adjusted exploration-exploitation method is adopted. Provided numerical results show that the proposed algorithm can achieve a higher average success rate and lower average overflow than baseline methods.

\end{abstract}


\section{Introduction}
Over the past decade, moving computing, control, and data storage into the cloud has been an important trend in order to utilize much needed abundant computing resources to handle explosive traffic demands. However, this relocation introduces network delays that bring significant challenges related to meeting the latency requirements of critical applications. To overcome the disadvantages of the cloud, fog computing, which selectively moves computation, communication, control, and decision making close to the network edge where data is being generated, became inevitable in this era \cite{b0}. One of the key benefits of fog computing stems from its highly virtualized platform that offers computing capacities allowing various applications to run anywhere. Hence, fog computing resolves problems of cloud-only solutions for applications that require a real-time response with low latency, e.g., mission-critical applications \cite{b1,b2}. Given the substantial benefits that can be drawn from this technology, fog computing is expected to play a crucial role in IoT, 5G, and other advanced distributed and connected systems \cite{b3,b4,b5,b6}. 

In fog networks, where fog nodes and cloud data centers present heterogeneous resources (e.g., computational, bandwidth, and memory), service tasks are classified according to various performance requirements and heterogeneous resource configurations. In contrast to the cloud server, the computing capacity of fog nodes is usually limited and in-homogeneous. Thus, computation-intensive tasks often exhibit poor performance when they are processed by fog nodes with extremely limited resource capacities \cite{b11, b11-1}. In this context, offloading and distributing tasks over the network while guaranteeing the Quality-of-Service (QoS) requirements of the users, can be particularly useful. Considering the fact that fog nodes are located relatively close to each other, offloading from an originally requested fog node to an affordable neighbor node with available resources can be an attainable solution even for delay-critical services. Moreover, it is critical for the fog nodes and cloud to be able to cope with heterogeneous tasks when deciding which service tasks should be deployed and where. Hence, both the fog and cloud should complement each other in a distributive way to fulfill service needs. To this end, the hierarchical fog architecture was introduced to support a better distribution of the computing resources with an elastic and flexible placement of resources \cite{b11-2}.

On the other hand, as wireless services and applications become more sophisticated and intelligent, there is a pressing need for an efficient management of the execution of increasingly complex tasks based on the application requirements. Specifically, the selection of suitable nodes and proper resource assignments are critical in fog networks, where various types of applications are simultaneously running over the same network \cite{b9-2, b9-3}. The problem is deteriorated due to the highly volatile service demands of end-users and uncertainties associated with resource availability at the fog nodes. When fog nodes handle significantly different traffic demands, the imbalance between fog nodes can lead to inefficient resource usage and inequitable QoS \cite{b9-4}. Furthermore, each node cannot acquire full knowledge of the other nodes due to a non-trivial amount of signaling overhead and communication latency. Therefore, how to distribute the computing resources optimally throughout the network and design algorithms based on local knowledge that can derive globally emergent system characteristics such as agility, efficiency, and reliability are the central questions that lead this paper \cite{b9-5}.

\subsection{Related work}
Recently, significant efforts have been centered on fog computing implementations to tackle the limitations of traditional cloud platforms. Specifically, many approaches have been proposed in the literature to enhance task offloading and resource allocation problems for fog networks.
Yousefpour \textit{et al.} \cite{b12} introduce a delay-minimizing offloading policy for fog nodes in IoT-fog-cloud application scenarios, where the policy considers not only the length of the queue but also different request types that vary in processing times. Following this, it determines whether or not to offload the selected tasks as the estimated waiting time of fog node is greater than an acceptance threshold; it will offload the request to its best neighbor fog node. 
Zhang \textit{et al.} \cite{b13} formulate the resource allocation problem between fog nodes, data service operators, and data service subscribers. First, service subscribers purchase the optimal number of computing resource blocks from service operators with a Stackelberg game. Each subscriber competes for the required computing resource blocks owned by the nearby fog nodes. With a many-to-many matching game between service operators and fog nodes, each operator determines its fog nodes that have computing resources to sell. With another many-to-many matching framework, resource blocks of fog nodes are allocated to the service subscribers.

Although some promising works have been dedicated to studying task offloading and resource allocation in fog computing and edge computing networks, it is necessary to jointly address the two issues to improve the overall performance. Wang \textit{et al.} \cite{b14} propose to jointly address computation offloading, resource allocation, and content caching in wireless cellular networks with mobile edge computing. First, they transform the original non-convex problem into a convex problem and prove the convexity of the transformed problem. Then, they decompose the problem and apply an alternating direction method of multipliers to solve it in an distributed and practical way.
Alameddine \textit{et al.} \cite{b15} address task offloading with joint resource allocation and scheduling specifically focused on delay-sensitive IoT services. They mathematically formulate the problem as a mixed-integer problem and present a decomposition approach to achieve faster run times while providing the optimal solution.
For heterogeneous real-time tasks, the authors in \cite{b16} propose task offloading and resource allocation problems in a three-tier fog system with a parallel virtual queueing model. They apply an adaptive queuing weight resource allocation policy based on the Lyapunov function. Moreover, they propose multi-objective sorting policies in terms of both the laxity and execution times of the task to achieve a trade-off between the throughput and task completion ratio optimization.

However, the computation offloading and resource allocation designs \cite{b12,b14,b15,b16} are mostly based on one-shot optimization and may not be able to achieve a long-term stable performance in dynamic situations. And since most optimization problems that arise in network resource management are non-convex and NP-hard, all these algorithms generally impose restrictions on the network to simplify non-trivial mathematical equations \cite{convex}. Nevertheless, such assumptions would require a revision of the objective functions, or even the system models, that lead to these problem formulations in the first place.

Furthermore, there are related works using different meta-heuristic methods \cite{m1,m2,m3,m4}. S. K. Mishra \textit{et al.} \cite{m2} introduce the scheduling of service requests to virtual machines (VMs) with the minimum energy consumption at the fog servers. They formulate the service allocation algorithm for the heterogeneous fog server system using three meta-heuristic methods, namely particle swarm optimization (PSO), binary PSO, and bat algorithm. Moreover, the authors in \cite{m4} introduce a new evolutionary algorithm (EA) that is combined with a PSO and genetic algorithm (GA) for the joint design of the computation offloading and fog nodes provisioning.

However, in meta-heuristic algorithms, the memory required to maintain a population of candidate solutions becomes vast as the size of problems increases. Specifically, due to the larger search space in large-scale problems, almost every state encountered will never have been seen before, which makes it impossible to converge in limited time steps. In that respect, as the system model becomes more complex, meta-heuristic methods can no longer be applied. In this context, to make sensible decisions in such large search spaces, it is necessary to generalize from previous encounters with different states that are in some sense similar to the current one.

In order to cope with an unprecedented level of complexity as we consider many parameters to accurately model the system, embedding versatile machine intelligence into future wireless networks is drawing unparalleled research interest \cite{r1,r2}. A lot of recent works try to address the resource allocation problem in IoT networks by using supervised machine learning, i.e., Support Vector Machines (SVMs), Recurrent Neural Networks (RNNs), Convolutional Neural Networks (CNNs), etc \cite{r3}. Nevertheless, supervised learning is learning from a fixed data set. Thus the algorithm does not directly interact with the environment where it operates, which is not adequate to dynamically provision the on-demand resources, especially for highly volatile IoT application demands. Moreover, in the context of resource management for IoT networks, the lack of sufficient labeled data is another factor that hinders the practicality of supervised learning-based algorithms. On the other hand, a different machine learning technique that does not fall in the category of supervised and unsupervised learning is reinforcement learning (RL) \cite{b20}. One of the key features of reinforcement learning is that it explicitly considers the problem of a goal-directed algorithm interacting with an uncertain environment. Therefore, RL-based algorithms can continually adapt as the environment changes without needing explicit system models. To tackle the curse of dimensionality of RL, deep reinforcement learning (DRL) was recently introduced \cite{b26}. DRL embraces deep neural networks to train the learning process, thereby improving the learning speed and the performance of RL-based algorithms. Therefore, a DRL can provide efficient solutions for future IoT networks \cite{r3-1}.

The authors in \cite{r3-2} introduce an optimal computation offloading policy for mobile edge computing (MEC) in an ultra dense system based on a deep Q-network (DQN) without prior knowledge of the dynamic statistics. Pan \textit{et al.} \cite{r3-3} study the dependency-aware task offloading decision in MEC based on Q-learning aiming to minimize the execution time for mobile services with limited battery power consumption.
Huynh \textit{et al.} \cite{r8} develop an optimal and fast resource slicing framework based on a semi-Markov decision process (MDP) that jointly allocates the computing, storage, and radio resources of the network provider to different slice requests. To further enhance the performance, they propose a deep Q-learning approach with a deep dueling neural network, which can improve the training process and outperform all other reinforcement learning techniques in managing network slicing. Chen \textit{et al.} \cite{r9} consider a software-defined radio access network where multiple service providers (SPs) compete to acquire channel access for their subscribed mobile users, thereby each mobile user proceeds to offload tasks and schedule queued packets over the assigned channel. They transform the stochastic game between non-cooperative SPs into an abstract stochastic game and propose a linear decomposition approach to simplify decision making. Also, a DRL algorithm is leveraged to tackle the huge state space. Sun \textit{et al.} \cite{r10} propose a DRL based joint communication mode selection and resource management approach with the objective of minimizing the network power consumption. This approach can help the network controller learn the environment from collected data and make fast and intelligent decisions to reduce power consumption.
Moreover, the tremendous growth in data traffic over next-generation networks can be substantially reduced via caching, which proactively stores reusable contents in geographically distributed memory storage \cite{r5,r4,r7,r6}. The authors in \cite{r4} study the joint cache and radio resource optimization on different timescales in fog access networks. The optimization problem is modeled as a Stackelberg game. To solve the problem, single-agent RL and multi-agent RL are utilized and rigorously analyzed. Meanwhile, the authors in \cite{r6} exploit the power allocation problem in non-orthogonal multiple access for a system with cache-enabled users. They propose a DRL based scheme, which responds quickly upon requests from users as well as allows all users to share the full bandwidth. Also, they show that applying iterative optimization algorithms is not suitable for satisfying a short-response requirement from the base station to users.

\subsection{Contributions}
This paper focuses on resource management in a fog system with the aim of guaranteeing the specific quality of service of each task as well as maximizing the resource utilization by cooperating between fog computing nodes. To address this problem, we design a joint heterogeneous task offloading and resource allocation algorithm whose goal is to maximize the processing tasks completed within their delay time limits. More precisely, we consider an independent multi-agent decision-making problem that is cooperative and partially observable. To solve this problem, we propose a deep recurrent Q-network (DRQN) based learning algorithm, namely deep Q-learning combined with a recurrent layer. The DRQN-based algorithm aims to resolve partially observable environments by maintaining internal states. In particular, to guarantee the convergence and accuracy of the neural network, the proposed DRQN-based algorithm adopts an adjusted exploration-exploitation scheduling method, which efficiently avoids the exploitation of incorrect actions as the learning progresses. To our best knowledge, this is the first work that introduces DRQN to solve the joint task offloading and resource allocation problems in fog computing networks.
The key contributions of this paper are summarized as follows.
\begin{itemize}
    \item The proposed algorithm considers two-levels of heterogeneity of service tasks in terms of QoS requirements and resource demand characteristics. In real IoT scenarios, various service tasks demanding different resource sizes can require different service performances. In order to consider these heterogeneities, we propose a fog network slicing model that manages different types of tasks separately and partitions physical resources to each slice.
    \item Regarding the feedback and memory overhead, we consider cooperative scenarios where the independent multi-fog nodes perceive a common reward that is associated with each joint action while estimating the value of their individual actions solely based on the rewards that they receive for their actions. Therefore, this reduces the feedback and memory overheads considerably compared to joint-action learning schemes where the fog nodes require the reward, observation, and action sets of others.
    \item To deal with partial observability, we apply a DRQN approach to approximate the optimal value functions. The DRQN-based algorithm can tackle partial observability issues by enabling the agent to perform the temporal integration of observations. This solution is more robust than DQN and deep convolutional Q-network (DCQN)-based methods in ways that the neural network with a recurrent layer can learn its output depending on the temporal pattern of observations by maintaining a hidden state, and thus it can keep internal states and aggregate observations. Moreover, to guarantee the convergence and accuracy of the neural network, an adjusted exploration-exploitation method is adopted.
    \item Numerical experiments using Tensorflow are presented to support the model and the proposed algorithm. The proposed DRQN-based algorithm requires much less memory and computation than the conventional Q-learning and meta-heuristic algorithms which are impractical for solving the problem considered in this paper. Particularly, the proposed DRQN-based algorithm is compared to the DQN and DCQN approaches where it is demonstrated that the performance in terms of average success rate, average overflow rate, and task delay can be significantly enhanced by using the proposed DRQN-based algorithm.
\end{itemize}

\subsection{Organizations}
The remainder of this article is organized as follows: in section \ref{sectiontwo}, the system description is presented. The formulation of the offloading and resource allocation problem as a partially observable MDP (POMDP) is detailed in Section \ref{sectionthree}. In section \ref{sectionfour}, we propose the cooperative decision-making problem between independent multi-nodes and derive a deep reinforcement learning scheme to solve the problem formulated in Section \ref{sectionthree}. Simulation results are presented in Section \ref{sectionfive}. Finally, Section \ref{sectionsix} concludes this paper and provides insight on possible future work.

\section{System description}\label{sectiontwo}
In this section, we introduce a three-layer fog network system model that supports the integration of different services while serving the best of each dissimilar service characteristics, such as CPU processing density and delay requirements, through a hierarchical model. The time horizon is divided into decision epochs of equal durations (in millisecond) and indexed by an integer $t\in{\mathbb{N_+}}$.  

\begin{figure}[t]
	\centering
	\includegraphics[width=1\linewidth]{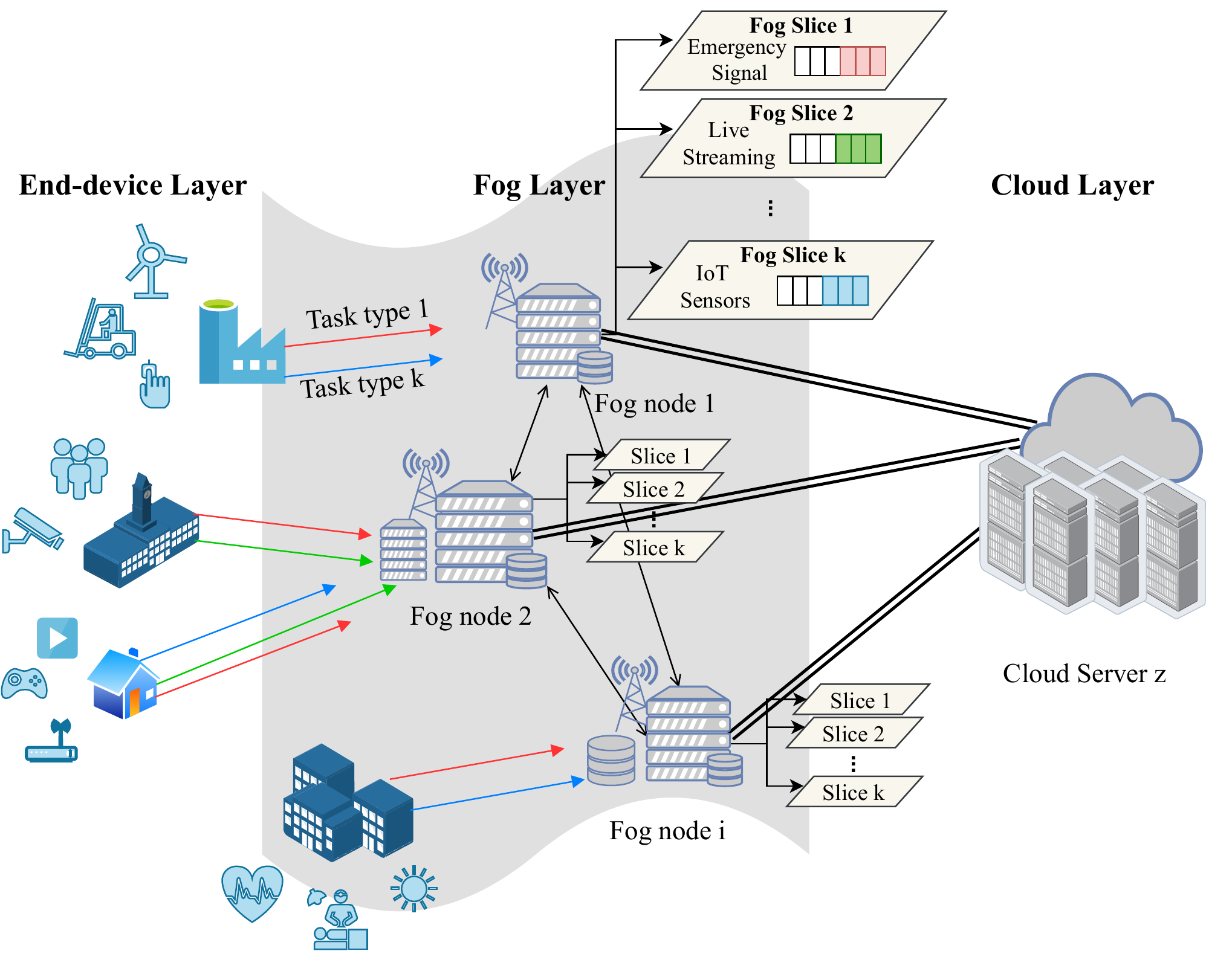}
	\caption{Three-layer fog network system.}
	\label{figure1}
\end{figure}
\begin{table}[]
\caption{List of Notations}
	\label{table0}
	\centering
    \begin{tabular}{l|l}
    \hline
    Symbol & \multicolumn{1}{|c}{Definition} \\ \hline
    \multicolumn{1}{c|}{$I$} & The set of fog nodes  \\ \hline
    \multicolumn{1}{c|}{$Z$} & The set of cloud servers \\ \hline
    \multicolumn{1}{c|}{$K_i$} & The set of fog slices at fog node $i$ \\ \hline
    \multicolumn{1}{c|}{$T_k$} & The packet size of the task of slice $k$ \\ \hline
    \multicolumn{1}{c|}{$D^{max}_k$} & The maximum delay budget of the task of slice $k$ \\ \hline
    \multicolumn{1}{c|}{$\lambda_{i,k}$} & The task of slice $k$ arrival rate for the fog node $i$ \\ \hline
    \multicolumn{1}{c|}{$a_{i,k}$} & \begin{tabular}[c]{@{}l@{}}The boolean variable whether the task of slice $k$ \\ arrives at fog node $i$ or not\end{tabular} \\ \hline
    \multicolumn{1}{c|}{$b_{i,k}$} & The number of tasks in the buffer of slice $k$ at fog node $i$ \\ \hline
    \multicolumn{1}{c|}{$b^e_{i,k}$} & \begin{tabular}[c]{@{}l@{}}The number of tasks are allocated resources for processing\\ among all the tasks in the buffer of slice $k$ at fog node $i$ \end{tabular} \\ \hline
    \multicolumn{1}{c|}{$L^c_k$} & CPU processing density demanded for the task of slice $k$\\ \hline
    \multicolumn{1}{c|}{$L^m_k$} & Memory size demanded for the task of slice $k$\\ \hline
    \multicolumn{1}{c|}{$U^c_i$} & Total CPU resource capacity of fog node $i$ \\ \hline
    \multicolumn{1}{c|}{$U^m_i$} & Total memory resource capacity of fog node $i$ \\ \hline
    \multicolumn{1}{c|}{$U^c_z$} & Total CPU resource capacity of cloud server $z$ \\ \hline
    \multicolumn{1}{c|}{$U^m_z$} & Total memory resource capacity of cloud server $z$ \\ \hline
    \multicolumn{1}{c|}{$\eta^c_i$} & The allocation unit of CPU resource at fog node $i$ \\ \hline
    \multicolumn{1}{c|}{$\eta^m_i$} & The allocation unit of memory resource at fog node $i$ \\ \hline
    \multicolumn{1}{c|}{$\eta^c_z$} & The allocation unit of CPU resource at cloud server $z$ \\ \hline
    \multicolumn{1}{c|}{$\eta^m_z$} & The allocation unit of memory resource at cloud server $z$ \\ \hline
    \multicolumn{1}{c|}{$BW_i$} & The transmission bandwidth of fog node $i$ \\ \hline
    \multicolumn{1}{c|}{$P_i$} & The transmission power of fog node $i$ \\ \hline
    \multicolumn{1}{c|}{$\beta_1, \beta_2$} & The path loss constant and exponent \\ \hline
    \multicolumn{1}{c|}{$r^c_i$} & The available CPU resource units at fog node $i$ \\ \hline
    \multicolumn{1}{c|}{$r^m_i$} & The available memory resource units at fog node $i$ \\ \hline
    \multicolumn{1}{c|}{$f_{i,k}$} & \begin{tabular}[c]{@{}l@{}} The offloading decision by fog node $i$ where \\ the task of slice $k$ will be processed \end{tabular} \\ \hline
    \multicolumn{1}{c|}{$w_{i,k}$} & \begin{tabular}[c]{@{}l@{}} The resource allocation decision by fog node $i$ how many \\ tasks of slice $k$ will be allocated resources for processing \end{tabular} \\ \hline
    \multicolumn{1}{c|}{$\psi_i$} & The local reward observed by fog node $i$ \\ \hline
    \end{tabular}
\end{table}
\subsection{Three-layer fog network system}
To improve scalability and resource utility in fog networks, a three-layer hierarchy is the most considered architecture \cite{b4,b11-2}. A three-layer fog network consists of an end-device layer, a fog layer, and a cloud layer. The end-device layer includes various types of devices, known as unintelligent devices, that are only producing data and are not equipped with processing capacity. Therefore, devices can request the nearest fog node to run their tasks on behalf of them. These tasks can be classified into types according to two characteristics, the task performance requirement (also called QoS) and the heterogeneous resource configuration (e.g., different CPU and memory configurations).  

The fog layer is composed of multiple fog nodes $i\in \mathcal{I}$. As illustrated in Fig. \ref{figure1}, we consider the fog layer where the physical network infrastructure is split into multiple virtual networks to offer heterogeneous service requests for different types of end-device segments, known as network slicing \cite{b9}. With network slicing technology, fog nodes can set up customized slices to guarantee specific latency and resource requirements by the supported services. Fog nodes are formed by at least one or more physical devices with high processing capabilities, which are aggregated as one single logical entity able to seamlessly execute distributed services as if it was on a single device. The shared physical resources (e.g., bandwidth, CPU, and memory) on fog nodes are partitioned into fog slices to enable running the network functions that meet certain required slice characteristics (e.g., ultra-low-latency, ultra-reliability, etc).

Finally, the cloud layer includes various servers that are capable of sufficient storage and computational resources but are physically remote from end-devices. In our architecture, a fog network comprises of a single cloud server $z\in \mathcal{Z}$ that interacts with all fog nodes.
\subsection{SDN-based fog nodes and inter-SDN communications}
To provide more distributed and scalable management, fog nodes make use of software-defined networking techniques where the control plane is capable of decision-making while the data plane simply serves forwarding and computing tasks \cite{b6-1, b6-2,b8-1}. Furthermore, many different applications are operated concurrently in the SDN application plane. Besides, as individual SDN controllers are located in separate fog nodes, we apply the concept of inter-SDN communications, which interconnects controllers to share information and coordinate their decisions \cite{b17}. It is noted that the need for inter-SDN communications is increased as the explosive increase in task demands of end devices is requiring networks formed by more than one SDN controller \cite{b17-1}. In our system model, each fog node defines, deploys, and adapts independent decision-making in its separate SDN controller, and communication between the SDN controllers of the fog nodes aims to exchange feedback information required by the independent decision-making processes. Details about the information exchange is provided in section \ref{sectionthree}.

\subsection{Fog slices based on heterogeneous task models}
We consider a fog network with fog nodes deploying the same set of logical fog slices for different task types over separate infrastructures (i.e. $\mathcal{K}_i=\mathcal{K}, \forall i\in \mathcal{I}$). Let $k\in \mathcal{K}_i$ be the set of slices available in the $i^{th}$ node where each slice processes a specific type of tasks in a separate buffer. The task demanded from the end devices by slice $k$ has a size of $T_k$ bits. In this context, since the time slot duration is relatively small, at most one task arrives at each slice of the fog node within a time slot. At the beginning of each time slot $t$, let $a_{i,k}(t)$ be the arrived task, where $a_{i,k}(t)= 1$ if a task of slice $k$ arrives at fog node $i$, otherwise $a_{i,k}(t)= 0$. Hence, the probability that a new task of slice $k$ arrives at fog node $i$ within time slot $t$ follows a Bernoulli distribution with parameter $\lambda_{i,k}$, $\mathbb{P}(a_{i,k}= 1)=\lambda_{i,k}$. The number of tasks in a buffer for slice $k$ at fog node $i$ and time slot $t$ is $b_{i,k}(t)$, of which $b^e_{i,k}(t)$ tasks are in progress in time slot $t$. Meanwhile, the maximum buffer size is $\overline{b}_{i,k}$.

Classified tasks in each slice have specific QoS requirements as well as different resource configurations. We assume that tasks delivered from end-devices are classified to the corresponding slices regarding their characteristics without manual intervention since classifying tasks to predict the type of application \cite{b18} are outside the scope of this paper. In terms of QoS, we categorize tasks into three classes: 1) delay-critical class (e.g., self-driving cars, live-streaming), 2) delay-sensitive class (e.g., augmented reality/virtual reality (AR/VR), smartphone applications) and 3) delay-tolerant class (e.g., IoT sensors). The priority of tasks is determined in a way that provides maximum reliability within an acceptable delay, proper to each slice.

On the other hand, with regard to resource configurations, tasks of each slice demand two types of resources (i.e. CPU and memory). To process a task of slice $k$, we denote the CPU processing density (in cycles/bit) and the memory (in Mbyte) demands as $L^c_k$ and $L^m_k$, respectively. Therefore, although tasks require the same QoS demands, the resource demands can be dissimilar \cite{b18-1}. One use case example of the delay-critical class is a live sport-streaming application requiring high-throughput, while another from the same class would be an emergency signal for self-driving cars which doesn't necessarily require high-throughputs. Thus, they are processed through different slices.

Furthermore, the total resource capacities (i.e. CPU and memory) of fog node $i$ and of the cloud server $z$ are $U_i=(U^c_i, U^m_i)$, $\forall i\in \mathcal{I}$ and $U_{z} =(U^c_z, U^m_z)$, respectively, where the superscript $c$ and $m$ indicate CPU speed (in cycles/$\Delta t$) and memory size (in Mbyte), respectively. We assume that the cloud server is much more computationally powerful than its associated fog nodes, i.e., $U^c_z\gg U^c_i$, $\forall i\in \mathcal{I}$, and provides limitless storage, i.e., $U^m_z\sim\infty$. The fog nodes and the cloud server can allocate their resource on a resource unit basis. Hence, the total amount of resource units, which can be allocated by fog node $i$ to all slices, can be computed as $N_i=(N^c_i, N^m_i)=\left(\lfloor{\frac{U^c_i}{\eta^c_i}}\rfloor, \lfloor{\frac{U^m_i}{\eta^m_i}}\rfloor\right)$ where $\eta^c_i$ and $\eta^m_i$ stand for the number of allocated units of computing and memory resources at fog node $i$, respectively, and $\lfloor\cdot\rfloor$ denotes the floor function. Likewise, $\eta^c_z$ and $\eta^m_z$ indicate the number of allocated units of computing and memory resources at cloud server $z$ and the total number of resource units of cloud server is unlimited.

Thus, for a given node $i$ at time $t$, the occupied resource units of all slices can be calculated as
\begin{equation} \label{one}
    G_i\left(t\right)=\left(g^c_i, g^m_j\right)=\left(\sum_{k}b^e_{i,k}(t), \sum_{k}b^e_{i,k}(t)\cdot\lceil{\frac{L^m_k}{\eta^m_i}}\rceil\right),
\end{equation}
where $\lceil\cdot\rceil$ is the ceiling function since a minimum of memory units greater than or equal to $\frac{L^m_k}{\eta^m_i}$ must be allocated to execute the task of slice $k$. 
At every time slot $t$, fog nodes only monitor their own available resources which correspond to the total resources minus the sum of resources being allocated to tasks of all slices. Hence, the available resource units at fog node $i$ and time $t$ can be measured as
\begin{equation} \label{two}
    R_i\left(t\right)=\left(r^c_i, r^m_j\right)=\left(N^c_i-g^c_i, N^m_i-g^m_i\right),
\end{equation}
where $r^c_i+g^c_i\le N^c_i$ and $r^m_i+g^m_i\le N^m_i$. Once the task processing is completed during the time slot, the finished task will be eliminated from the buffer and the resource allocated to this task will return to the available resource pool in the next time slot. 

\subsection{Calculation of task latency}
In our architecture, at the beginning of each time slot $t$, fog nodes use an independent offloading policy to decide whether they will process arrived tasks locally or offload them to another node between neighboring fog nodes and the cloud server. Furthermore, decisions on resource allocation are simultaneously made by fog nodes with regard to their own resources through separate allocation policies.

We define a task latency to enable different delay constraints for tasks, thereby minimizing timeout failures that result from high transmission latency from offloading to a remote node or long waiting delays due to insufficient resource capacities at the processing node. Formally, the task latency can be denoted as the sum of the transmission delay, waiting delay, and processing delay. We assume that the fog nodes have information regarding the distance to neighboring fog nodes in the fog network as well as to the cloud server. To model the transmission delay of offloading, the tasks are transmitted to the selected node over a wireless channel. Then the transmission delay for fog node $i$ to forward the task of slice $k$ can be defined as
\begin{equation} \label{three}
	D^s_{i,j,k,n}(t)= 
\begin{cases}
    \frac{T_k}{\nu_{i,j,n}(t)},& \text{if } i\neq j \\
    0,              & \text{otherwise,}
\end{cases}
\end{equation}
where $j\in \{\mathcal{I}, \mathcal{Z}\}$ if the selected node is a fog or cloud node, respectively, and $n \in \{1,2,..K\}$ indicates the total number of tasks offloaded by fog node $i$ at time slot $t$. Moreover, $\nu_{i,j,n}(t)$ represents the transmission rate from fog node $i$ to the selected node $j$, which is given by \cite{b11}: 
\begin{equation} \label{four}
	\nu_{i,j,n}(t)=BW_{i,n}(t)\cdot\log\left(1+\frac{\beta_1 {d_{i,j}}^{-\beta_2}\cdot P_i}{BW_{i,n}(t)\cdot\sigma^2}\right),
\end{equation}
where $d_{i,j}$, $\beta_1$, and $\beta_2$ are the distance between two nodes, the path loss constant, and the path loss exponent, respectively. The variable $P_i$ denotes the transmission power of fog node $i$ and $\sigma^2$ is the noise power spectral density. Additionally, the bandwidth is given by $BW_{i,n}(t)=\frac{BW_i}{n}$, which means that the total bandwidth of the fog node $BW_i$ is equally shared by $n$ tasks. For example, when a fog node $i$ offloads a total of two different tasks during a time slot, each task is transmitted with $\frac{BW_i}{2}$ in separate ways. When $i=j$, a fog node $i$ processes this task locally, thus there is no transmission delay. Moreover, in most cases, the size of task after processing is small, thus the transmission delay of the result after processing can be ignored.

Next, when the slice $k$ task arrives in the corresponding buffer at node $j$, the waiting delay $D^q_{i,j,k}(t)$ can be calculated as
\begin{equation} \label{five}
	D^q_{i,j,k}(t)=
	\begin{cases}
    \frac{b_{j,k}(t)}{\mu_{j,k}(t)},& \text{if } j\in I \\
    0,              & \text{otherwise,}
    \end{cases}
\end{equation}
where $b_{j,k}(t)$ is the number of tasks previously existing in a buffer and $\mu_{j,k}(t)$ is the service rate (i.e., the rate of tasks leaving a buffer). However, since a fog node does not have prior information about the buffer status of the other nodes when offloaded tasks arrive in its buffer and given that the service rate varies depending on the resource scheduling process, the waiting delay cannot be calculated in advance. On the other hand, we assume that the waiting time at the buffer of the cloud can be disregarded because the cloud is equipped with a larger number of cores than the fog node. This indicates that the cloud initiates the computation for the received tasks without queueing delay.

When a task is computed by fog nodes, the processing delay $D^p_{i,j,k}(t)$ can be denoted as 
\begin{equation} \label{six}
	D^p_{i,j,k}(t)=
	\begin{cases}
    \frac{T_k\cdot L^c_k}{\eta^c_j},& \text{if } j \in I 
    \\[8pt]
    \frac{T_k\cdot L^c_k}{\eta^c_z},& \text{otherwise, }
    \end{cases}
\end{equation}
where $T_k\cdot L^c_k$ refers to the number of CPU cycles required to complete the execution of a task of slice $k$. When the task is offloaded to a fog node $ j \in \mathcal{I}$, the task of slice $k$ is executed by a fog node $j$ with the CPU speed $\eta^c_j$. Likewise, when the task is offloaded to the cloud server $z$, the processing delay is formulated in the bottom equation of (6) where $\eta^c_z$ is the CPU speed of cloud server $z$. Thus, the processing delay is dependent on both the resource configuration of the task and the amount of allocated resources.

In essence, if a slice $k$ task is offloaded from a fog node $i$ to a neighboring fog node $j\neq i$, the latency is obtained as 
\begin{equation} \label{seven}
    \begin{split}
    D_{i,j,k,n}(t)&= D^s_{i,j,k,n}(t)+D^q_{i,j,k}(t)+D^p_{i,j,k}(t) \\
    & = \frac{T_k}{\nu_{i,j,n}(t)}+\frac{b_{j,k}(t)}{\mu_{j,k}(t)}+\frac{T_k\cdot L^c_k}{\eta^c_j}.
    \end{split}
\end{equation}
If a slice $k$ task is computed locally by a fog node $i$, the latency becomes 
\begin{equation} \label{eight}
	\begin{split}
    D_{i,i,k,n}(t)&= D^q_{i,i,k}(t)+D^p_{i,i,k}(t) \\
    & = \frac{b_{i,k}(t)}{\mu_{i,k}(t)}+\frac{T_k\cdot L^c_k}{\eta^c_i}.
    \end{split}
\end{equation} 
Finally, if a slice $k$ task is offloaded from a fog node $i$ to the cloud server $z$, the latency is
\begin{equation} \label{nine}
	\begin{split}
    D_{i,z,k,n}(t)&= D^s_{i,z,k,n}(t)+D^p_{i,z,k}(t) \\
    & = \frac{T_k}{\nu_{i,z,n}(t)}+\frac{T_k\cdot L^c_k}{\eta^c_z}.
    \end{split}
\end{equation}

\section{Problem formulation}\label{sectionthree}
In this section, we define the problem of heterogeneous task offloading and resource allocation in a system with multiple fog nodes as a POMDP across the time horizon.

\subsection{Partially observable MDP based problem formulation}\label{sectionthreea}
The main goal of the system is to make an optimal offloading and resource allocation decision at each node with the objective of maximizing the successfully processed tasks while guaranteeing the corresponding delay constraint of each task. Therefore, the joint offloading and resource allocation decision is achieved by finding proper processing nodes for the tasks and an optimal allocation of the node’s resources to all individual slices. We assume that the joint offloading and resource allocation decisions from all fog nodes are made simultaneously at every time slot $t$. 
To this end, each node repeatedly observes its own system states at the beginning of the time slot. The local observation by fog node $i$ is defined as 
\begin{equation} \label{ten}
	O_i(t)=\Big(A_i(t), B_i(t), B^e_i(t), R_i(t)\Big),
\end{equation}
where $A_i(t)=(a_{i,k}(t): k\in \mathcal{K}_i)$, $B_i(t)=(b_{i,k}(t): k\in \mathcal{K}_i)$, and $B^e_i(t)=(b^e_{i,k}(t): k\in \mathcal{K}_i)$ are the set of arrived tasks, the number of tasks in the buffer, and the number of tasks in progress among $B_i(t)$ from all slices at the fog node $i\in \mathcal{I}$ and time $t$, respectively. Moreover, $R_i(t)=\Big(r^c_i(t),  r^m_i(t)\Big)$ is the available resource units at the fog node $i\in \mathcal{I}$ at time $t$.

Note that the underlying states of the system including the states of other fog nodes are not accessible by the fog node. Instead, only the aforementioned state set in (\ref{ten}) can be observed and thus the system becomes a POMDP. We suppose that the observations are limits to the measurement accuracy of the state but are enough to make usable state data for a POMDP system.

In the presence of uncertainties stemming from the task demands and resource availability at the fog nodes, we formulate the POMDP based problem across the time horizon as a stochastic game in which each node selects actions as a function of their local observation. In our model, a fog node’s offloading and resource allocation policy operates independently from the other nodes’ policies. Thus, each fog node does not have any prior information on the task demands, buffer status, and resource availability of the other fog nodes.
Accordingly, the actions are defined as
\begin{equation} \label{eleven}
	X_i(t)=\Big(X^f_i(t), X^w_i(t)\Big),
\end{equation}

\noindent where $X^f_i(t)=(f_{i,k}(t): k\in \mathcal{K}_i)$ and $X^w_i(t)=(w_{i,k}(t): k\in \mathcal{K}_i)$ denote the offloading decision and resource allocation decision, respectively. $f_{i,k}(t)\in\{0,1,...,I+1\}$ represents by whom the task will be processed, where $f_{i,k}(t)=0$ if the slice $k$ task doesn't arrive at the fog node $i$ at time $t$, $f_{i,k}(t)=i$ if the slice $k$ task arrives and will be computed locally, and $f_{i,k}(t)=j, j\in \mathcal{I}$ and $j\neq i$, if the slice $k$ task arrives and will be offloaded to another node $(f_{i,k}(t)=I+1$ implies that the fog node will offload this task to the cloud server). The resource allocation decision $w_{i,k}(t)\in\{0,1,...,\lfloor\frac{U^m_i}{L^m_k}\rfloor\}$ represents how many tasks will be initiated by being allocated resources where $\lfloor\frac{U^m_i}{L^m_k}\rfloor$ is the maximum number of tasks that can be simultaneously processed by the fog node $i$. For example, $w_{i,k}(t)=2$ indicates that, at time $t$, fog node $i$ allocates its resources to slice $k$ to execute two tasks that are not in progress in the buffer. Each node takes an action $X_i(t)$ only among the ones allowed in that observation, i.e., $X_i(t)\in\mathcal{X}_i(O(t))$.
We apply the following constraints for the offloading and resource allocation at time $t$, 
\begin{equation} \label{twelve}
	f_{i,k}(t)=0, \text{if }a_{i,k}(t)=0, \forall k\in \mathcal{K}_i, \forall i\in \mathcal{I},
\end{equation}
\begin{equation} \label{thirteen}
	w_{i,k}(t)\le (B_i(t)-B^e_i(t)), \forall k\in \mathcal{K}_i, \forall i\in \mathcal{I},
\end{equation}
\begin{equation} \label{fourteen}
	\sum_{k\in \mathcal{K}_i} w_{i,k}(k)\le r^c_i, \forall i\in \mathcal{I},
\end{equation}
\begin{equation} \label{fifteen}
	\sum_{k\in \mathcal{K}_i} w_{i,k}(k)\cdot\lceil{\frac{L^m_k}{\eta^m_i}}\rceil\le r^m_i, \forall i\in \mathcal{I}
\end{equation}
to ensure that the fog node cannot offload the task when it doesn't arrive by (\ref{twelve}), cannot allocate more than the number of tasks waiting for allocation by (\ref{thirteen}), and the sum of newly allocated resources cannot exceed the available resources by (\ref{fourteen}) and (\ref{fifteen}).

Given that each node is in state $O_i(t)$ and action $X_i(t)$ is chosen, a transition probability is given by (\ref{sixteen}), where $\mathbf{X}(t)=(X_i(t): i\in \mathcal{I})$ are the set of actions occurring at time $t$. From (\ref{sixteen}), $B_i^e(t+1)$ and $R_i(t+1)$ only depend on the action $X_i(t)$ of fog node $i$, while $A_i(t+1)$ is determined regardless of the action. Since one node's offloading decisions result in increasing others' buffers, the sequence of each node's buffer status $B_i(t)$ depends on the actions of all agents $\mathbf{X}(t)$. 
 \begin{figure*}[t]
    \begin{equation}\label{sixteen}
    \begin{aligned}
    \mathbb{P}\Big(O_i(t+1)|O_i(t), \mathbf{X}(t)\Big)=&  \mathbb{P}\Big(A_i(t+1)\Big)\times\mathbb{P}\Big(B_i(t+1)|B_i(t),\mathbf{X}(t)\Big)\\
    &\times\mathbb{P}\Big(B_i^e(t+1)|B_i^e(t),X_i^w(t)\Big)\times\mathbb{P}\Big(R_i(t+1)|R_i(t), X^w_i(t)\Big)
    \end{aligned}
    \end{equation}
    
    \begin{equation} \label{seventeen}
    \begin{aligned}
    \psi_i(O_i(t), \mathbf{X}(t))&=\frac{1}{K}\cdot\sum_{k\in \mathcal{K}}a_{i,k}(t)\cdot\Big((-1)^{\mathds{1}(D^{max}_k\le D_{i,k}(t))}-\xi_k\cdot\mathds{1}(b_{f_{i,k},k}(t+D^t_{i,k}(t))\ge\overline{b}_{f_{i,k},k})\Big)
    \end{aligned}
    \end{equation}
     \hrulefill
\end{figure*}

Based on the set of actions $\mathbf{X}(t)$ in local observation $O_i(t)$, we define the local reward in (\ref{seventeen}),
where $D_k^{max}$ is the maximum delay budget of the task in slice $k$ where the task is discarded if its processing is not completed within this budget.
The first term of the summation in (\ref{seventeen}) represents the success reward, a positive reward if the task is successfully completed and negative reward if timeout failure is encountered, which depends on both offloading decisions of arrived fog node $i$ and resource allocation decision of the processing fog node. The second term describes the overflow cost which defines whether the task is dropped because the slice buffer is already full, thus it is related to the buffer status of processing fog node $b_{f_{i,k},k}$. Moreover, $\xi_k$ is a constant weighting factor that balances the importance of the overflow failure for tasks of slice $k$.

\subsection{Cooperative games by independent learners}
Although each fog node’s main goal is to optimize its own service performance and its resource interests, the fog nodes must still coordinate on the resource flows between neighboring nodes in order to achieve a meaningful solution from an overall system perspective \cite{b21}. In addition, the service performance experienced by service tasks during the processing is determined by the offloading and the resource allocation decisions of all fog nodes. Therefore, our stochastic game, sometimes called Markov game, follows a cooperative network to maximize the common goal rather than a competitive game where each fog node has opposing goals \cite{b22, b23}.

More precisely, we apply cooperative scenarios between independent multi-fog nodes where the fog nodes  share their local rewards with others as feedback information. This decision-making problem implies that independent fog nodes perceive the common reward that is associated with each joint-action while estimating the value of their individual actions solely based on the rewards that they receive for their actions. Therefore, this reduces the feedback and memory overheads considerably compared to joint-action learning schemes where the fog nodes share their reward, observation, and action sets with others to maintain a model of the strategy of other agents.
As such, at each time step, each node executes an individual action, with the joint goal of maximizing the average rewards of all nodes which can be formally formulated as 
\begin{equation} \label{eighteen}
	\psi(t)=\sum_{i\in \mathcal{I}}{\psi_i\Big(O_i(t), \mathbf{X}(t)\Big)}.
\end{equation} 
Thus, each node's reward is drawn from the same distribution, reflecting the value assessment of all nodes \cite{b23-2}. Moreover, the convergence performance of joint-action learning schemes is not enhanced dramatically despite the availability of more information due to the exploration strategy \cite{b23-2}.
As detailed in section \ref{sectiontwo}, the reward feedback is transmitted through inter-SDN communications to the SDN controllers of all the fog nodes for the decision-making process. In summary, the decision-making process at each fog node is fully distributed for real-time task offloading and resource management while communications between SDN controllers aims to exchange less time-sensitive reward information.

\section{Learning the optimal offloading and resource allocation policies}\label{sectionfour}
In this section, we propose a Q-learning-based optimal policy solution to address the limitations of the traditional approaches and discuss deep recurrent Q-networks (DRQN) which can better approximate actual Q-values from sequences of observations, leading to better policies in a partially observable environment.

\subsection{Optimal policy solution using Q-learning}
In the case where the system has access to transition probability functions and rewards for any state-action pair, the MDP can be solved through dynamic programming (DP) approaches to find the optimal control policy \cite{b24, b25}. However, in our cases, the system cannot precisely predict the transition probability distributions and rewards. To address this limitation, reinforcement learning is proposed in which the lack of information is solved by making observations from experience \cite{b20}. Among the different reinforcement learning techniques, Q-learning is used to find the optimal state-action value for any MDP without an underlying policy. Given the controlled system, the learning node $i$ repeatedly observes the current state $O^t_i$, takes action $X^t_i$ that incurs a transition, then it observes the new state $O^{t+1}_i$ and the reward $\psi^t_i$. From these observations, it can update its estimation of the Q-function for state $O^t_i$ and action $X^t_i$ as follows:
\begin{equation}\label{19}
    \begin{aligned}
    Q_i(O^t_i,X^t_i) \leftarrow &(1-\alpha)\cdot Q_i(O^t_i,X^t_i)\\
    &+\alpha\cdot[\psi^t_i+\gamma\max_{X'\in \mathbf{X}(O^{t+1}_i)}Q_i(O^{t+1}_i,X')],
    \end{aligned}
\end{equation}
where $\alpha$ is the \textit{learning rate} (0$<$$\alpha$$<$1), balancing the weight of what has already been learned with the weight of the new observation, and $\gamma$ is the \textit{discount factor} (0$<$$\gamma$$<$1).The most common action selection rule is the $\epsilon$-greedy algorithm that behaves greedily most of the time i.e., Greedy selection ($X^t\doteq\arg\max_{X'}Q(O^t,X')$) and explores other options by selecting a random action with a small probability $\epsilon$. This greedy selection and the $\epsilon$ probability of random selection are called exploitation and exploration, respectively. Non-optimal action selection can be uniform during exploration ($\epsilon$-greedy algorithm) or biased by the magnitudes of Q-values (such as Boltzmann exploration).

Moreover, we discuss the computational complexity of the Q-learning algorithm. The Q-algorithm requires storing a $|\mathcal{O}|\times|\mathcal{X}|$ size table of Q-values, i.e., $Q(O,X)$ for all $O\in \mathcal{O}$ and $X\in \mathcal{X}$. In our problem, the size of local observation spaces $|\mathcal{O}_i|$ and local action spaces $|\mathcal{X}_i|$ is calculated as $\prod_{k\in\mathcal{K}}\Big(2\times(1+\overline{b}_{i,k})^2\Big)\times(1+N^c_i)\times(1+N^m_i)$ and $(I+2)^K\times\prod_{k\in\mathcal{K}}\Big(1+\lfloor\frac{U^m_i}{L^m_k}\rfloor\Big)$. When $I=5$, $K=3$, $\overline{b}_{i,k}=5$, $N^c_i=5$, $N^m_i=5$, and $\lfloor\frac{U^m_i}{L^m_k}\rfloor=5$, one node $i$ has to update a total of $9.955\times10^{8}$ Q-function values, which makes it impossible for the conventional Q-learning process to converge within a limited number of time steps. This problem is even more pronounced in multi-agent scenarios, where the number of joint actions grows exponentially with the number of agents in the system.

\subsection{Convergence to equilibrium} $\pi^*_i$ of a node $i$ is the optimal policy to other nodes. Recall from fictitious play \cite{b23-3}, the exploration strategy is required to be asymptotically myopic to ensure that Nash equilibrium will be reached in multi-agent RL strategies. An action selection rule $\pi_i$ is said to be asymptotically myopic if the loss from agent $i$’s choice of actions at every given history $\pi_i$  goes to zero as $t$ proceeds\cite{b23-3}:
\begin{equation} \label{20}
	\psi(\pi^t_i)\nearrow\max\{\psi(X_i)|X_i\in\mathcal{X}_i(O_i)\},
\end{equation} 
as $t\to\infty$, where $\psi$ denotes the reward function. 
Therefore, the independent multi-agent Q-learning in cooperative systems will converge to equilibrium almost surely when the following conditions are satisfied \cite{b23-1}:
\begin{itemize}
    \item The learning rate $\alpha$ decreases over time such that $\sum^t\alpha=\infty \text{ and } \sum^t\alpha^2<\infty$.
    \item Each node visits every action infinitely often.
    \item The probability $\mathbb{P}_i^t(x)$ that node $i$ selects action $x$ is nonzero, $x\in\mathcal{X}(o)$.
    \item The exploration strategy of each node is exploitative such that $$\lim_{t\to\infty}\mathbb{P}_i^t(\pi^t_i)=0\text{ ,}$$ where $\pi^t_i$ is a random variable denoting a non-optimal action was taken based on estimated Q-values of node $i$ at time $t$. 
\end{itemize}
The first two conditions are required for convergence in Q-learning, while the third ensures that nodes explore with a positive probability at all times, which will ensure the second condition. Last but not least, the fourth condition guarantees that agents exploit their knowledge as the number of time steps increases. In fact, convergence of Q-learning does not depend on the exploration strategy used, which implies that there is no rule to choose actions as long as every action is visited infinitely often. However, effective exploration strategies will encourage long run optimal equilibrium \cite{b23-1}. To this end, we propose an adjusted exploration-exploitation method in the next subsection.

\subsection{Deep Q-learning with nonlinear transformation}
\begin{figure}[t]
	\centering
	\includegraphics[width=1\linewidth]{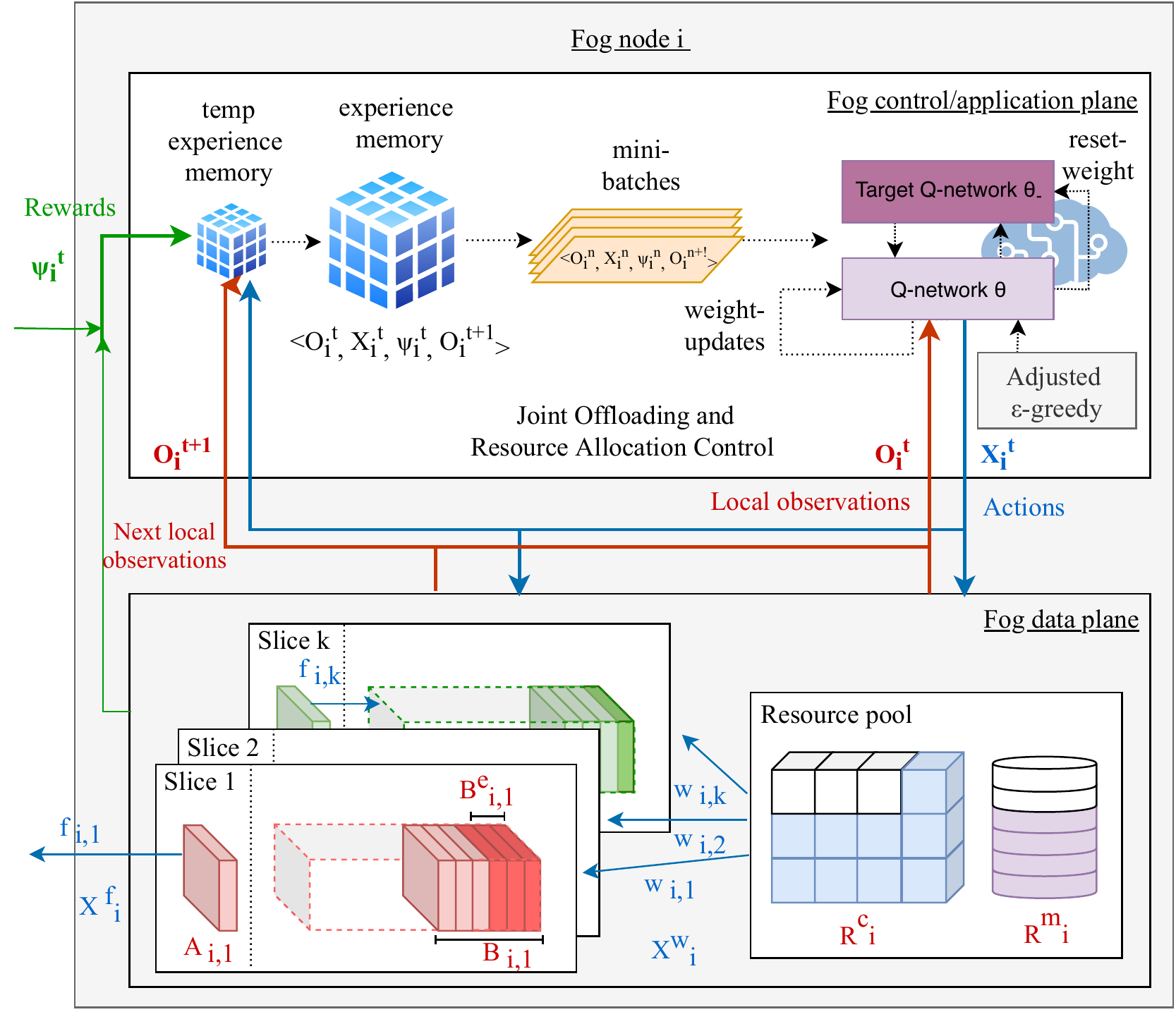}
	\caption{Application of a deep Q-network (DQN) to approximate the optimal joint offloading and resource allocation control policy of the SDN-based fog nodes.}
	\label{figure2}
\end{figure}
\begin{figure*}[t]
    \begin{equation}\label{twenty}
    \begin{aligned}
    \mathcal{L}_i\left({\theta}^t_i\right) = \mathbb{E}\left[\left(\psi^t+\gamma\max_{X'\in \mathcal{X}_i}Q_i\left(O_i^{t+1},X';{\theta}^{t-1}_i\right)-Q_i\left(O_i^t,X_i^t;{\theta}^t_i\right)\right)^2\right]
    \end{aligned}
    \end{equation}
    
    \begin{equation} \label{twentyone}
    \begin{aligned}
    \nabla_{{\theta}^t_i}\mathcal{L}_i\left({\theta}^t_i\right)=\mathbb{E}\Big[\Big(\psi^t+&\gamma\max_{X'\in \mathcal{X}_i}Q_i\left(O_i^{t+1},X';{\theta}^{t-1}_i\right)-Q_i\left(O_i^t,X_i^t;{\theta}^t_i\right)\Big)\cdot\nabla_{{\theta}^t_i} Q_i\left(O_i^t,X_i^t;{\theta}^t_i\right)\Big]
    \end{aligned}
    \end{equation}
     \hrulefill
\end{figure*}
To solve the scalability issues of Q-learning, we adopt Q-learning with a neural network, called deep Q-network (DQN). The DQN embraces the advantage of deep neural networks (DNNs) to train the learning process at each node $i\in\mathcal{I}$, thereby improving the learning speed and the performance of Q-learning algorithms.

The Q-network can be trained by iteratively adjusting the weights $\theta$ to minimize a sequence of the loss function, $\mathcal{L}_i\left({\theta}^t\right)$, where the loss function at time slot $t$ is defined in (18).
Precisely, given a transition $\langle O_i^t,X_i^t,\psi^t,O_i^{t+1}\rangle$, the weights $\theta^t_i$ of the Q-network of node $i\in\mathcal{I}$ are updated in a way that minimizes the squared error loss between the current predicted Q-value of $Q_i\left(O_i^t,X_i^t\right)$ and the target Q-value of $\left[\psi^t+\gamma\max_{X'\in \mathcal{X}_i}Q_i(O_i^{t+1},X')\right]$. The gradient of the loss function in (\ref{twenty}) with respect to the weights ${\theta}^t_i$ is given by (\ref{twentyone}).

Moreover, in the DQN algorithm, the experience replay technique is adopted as the training method to address the instability of the Q-network due to the use of non-linear approximation functions \cite{b26}. Hence, the transition experiences, $e^t_i=\langle O_i^t,X_i^t,\psi^t,O_i^{t+1}\rangle$ are stored into a replay buffer $\mathcal{M}_i=\{e^{t-\mathcal{D}_i}_i,…,e^t_i\}$, where $\mathcal{D}_i$ is the replay buffer capacity. Due to possible delays in the reward feedback between fog nodes, the past transition experiences may need to wait in the temporal replay buffer to combine the rewards (as in (\ref{eighteen})), where they are transferred to the replay buffer as soon as it is ready. At each time step, instead of the most recent transition $e^t_i$, a random mini-batch $\mathcal{N}_i$ of transitions from the replay memory is chosen to train the Q-network by node $i$. Furthermore, every $\mathcal{C}$ time steps, the network $Q_i$ is duplicated to obtain a target network $\hat{Q}_i$ which is used for generating the target Q-value for the following $\mathcal{C}$ updates to $Q_i$.

In addition, to guarantee the convergence and accuracy of the neural network, we adopt an adjusted exploration-exploitation scheduling method. At the beginning of the process, the agent with a normal $\epsilon$-greedy algorithm selects more random actions with a probability $\epsilon=\epsilon_{start}$ to encourage initial exploration. Then, the exploration rate is asymptotically decayed with $\epsilon_{decay}$ until it reaches a certain minimum value $\epsilon_{min}$ and is preserved until the last iteration. Since $\epsilon_{min}$ is a very small number, after this initial exploration phase, most decisions take the highest estimated value at the present iteration. However, this often leads to a sub-optimal policy due to exploiting bad estimates of the Q-value which were learned during the early iterations and insufficient exploration in large state-action space cases. To deal with this problem. the adjusted exploration-exploitation method allows the agent to shift back into exploratory mode every $\mathcal{R}^\epsilon$ time slots, where the starting exploration probability $\epsilon_{start}$ is decreased $\delta^\epsilon$ (0$<$$\delta^\epsilon$$<$1) times every update. Therefore, this method efficiently avoids the exploitation of incorrect actions by selecting better estimates of the Q-value as the learning progresses.
The optimal control policy learning implementation using the DQN algorithm is illustrated in Fig. \ref{figure2}.

\subsection{Deep-recurrent Q-learning for partial observability}
Another problem is that estimating a Q-value from an immediate observation in DQN can be arbitrarily wrong since the network states are partially observed and hinge upon multiple users \cite{b27}. Any system that requires a memory of more than an immediate observation will appear to be non-Markovian because the future system states depend on more than just the current input. This issue can be solved by allowing the agent to perform temporal integration of observations. The solution adopted in \cite{b26} stacks the last four observations in memory and feeds them to the convolutional neural network (called DCQN) instead of a single observation at a time. However, the DCQN takes in a fixed size vector as input, a stack of 4 observations in \cite{b26}, which limits its usage in situations that involve a sequence type input with no predetermined size.
In order to address this issue, we implement a DRQN which replaces the DCQN’s first fully connected layer by a recurrent layer. By utilizing a recurrent layer, the neural network will be able to learn its output depending on the temporal pattern of observations by maintaining a hidden state that it computes at every iteration. The recurrent layer can feed the hidden state back into itself, and thus it can maintain internal states and aggregate observations. 
\begin{figure*}
  \centering\includegraphics[width=1\linewidth]{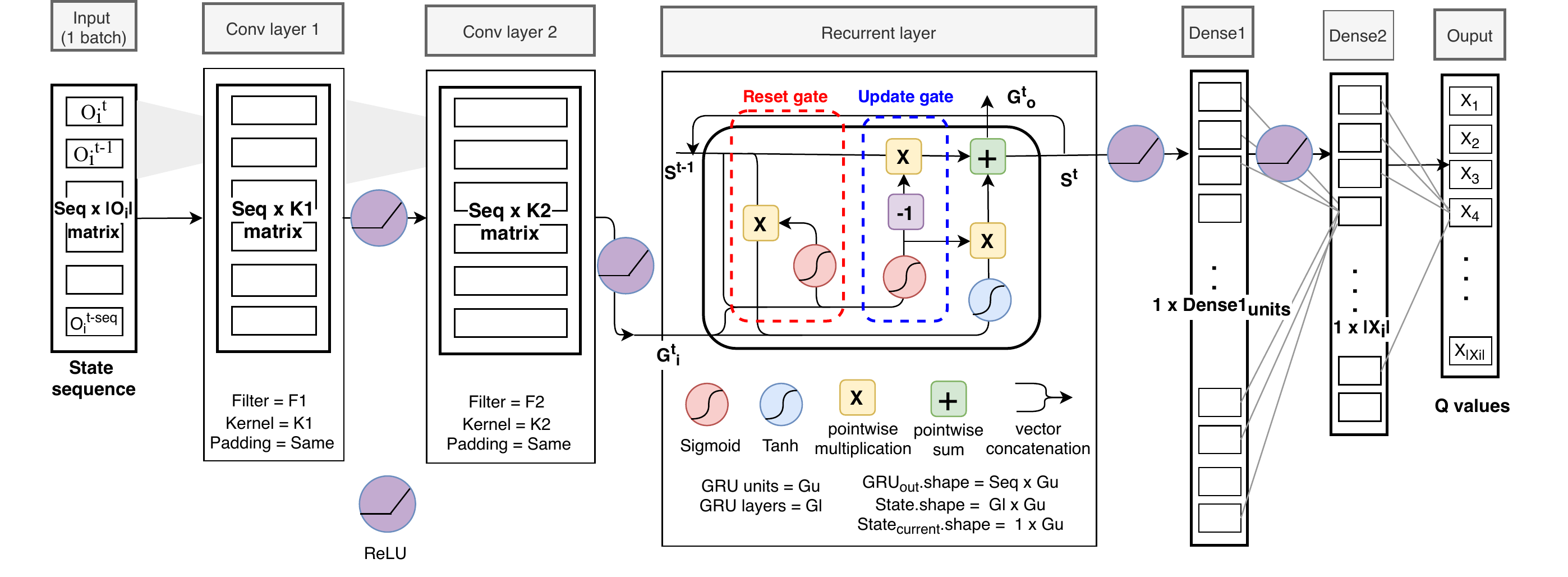}\par
  \caption{The proposed DRQN structure with GRU.}
  \label{DRQN}
\end{figure*}

However, during backpropagation, vanilla recurrent neural networks suffer from the vanishing gradient problem, which makes layers that get a small gradient value stop learning, and thus neural networks may forget important information from the beginning. To tackle this problem, we use Gated Recurrent Unit (GRU) for the recurrent layer. Similar to Long short-term memory (LSTM), GRU was introduced as a solution to the short-term memory of vanilla recurrent neural networks \cite{b28}. The main concept of GRUs is a gating mechanism, which can learn which information is relevant to keep or forget during training in the recurrent network. GRU has two gates (reset and update) and is known to be computationally more efficient and faster than LSTM which consists of three gates (forget, input, and output) and cell state, while its performance is comparable to LSTM \cite{b28, b29}.
The proposed DRQN structure is illustrated in Fig. \ref{DRQN}.
\begin{equation}
  \begin{split}
    G^t_r &=\sigma(W^s_{r}S^{t-1}+W^g_{r}G^t_{i}+\text{bias}_{r}), \\
    G^t_z &=\sigma(W^s_{z}S^{t-1}+W^g_{z}G^t_{i}+\text{bias}_{z}), \\
    \tilde{S}^t &=\tanh(W^{s}(G^t_r\odot S^{t-1})+W^{g}G^t_{i}+\text{bias}), \\
    S^t &=G^t_z\odot S^{t-1}+(1-G^t_z)\odot\tilde{S}^t,
  \end{split}
\end{equation}
where $G^t_r$ and $G^t_z$ are reset and update gates, respectively. With that, the recurrent network can learn how to use some of its units to selectively cancel out the irrelevant information and protect the state. Sigmoid and Tanh activation functions can make these decisions by filtering values between 0 and 1 for each state element.

Algorithm 1 details the procedure of the proposed learning algorithm at fog node $i$. The neural network takes the state sequence as an input to the first convolutional layer. Since the valid action space is dependent upon the current state value, we involve a step in the action selection that sets the probability of invalid actions to zero and re-normalizes the sum of the probabilities of the other actions to 1.

\begin{algorithm}
	\DontPrintSemicolon
	\LinesNumbered
	\textbf{Set} Initialize replay buffer $\mathcal{M}_i$ to capacity $\mathcal{D}_i$, state-action value function $Q_i$ with random weights $\theta_i$, target state-action function $\hat{Q_i}$ with weights $\theta_{i-}=\theta_i$.
	
	\While{(t $\le$ maximum iteration)}
	{
		Observe the arrival task $A_i^t$, buffer state $B_i^t, B^{e,t}_i$, resource status $R_i^t$ and combine them as $O_i^t$ and take $\hat{O}_i^t$ as an input to the DRQN network with parameter $\theta_i$, where $\hat{O}_i^t$ is a state sequence \;
		Calculate $\epsilon=\max[\exp{(-\epsilon_{decay}\cdot(t\mod{\mathcal{R}^\epsilon})+\log{\epsilon_{start}})}, \epsilon_{min}]$, and choose random action $X_i^t$ from valid action spaces $X(O_i^t)$ with probability $\epsilon$ otherwise select $X_i^t\doteq\arg\max_{X'}Q(O_i^t,X';\theta_i)$\;
		Execute action $X_i^t$; offload tasks according to $X_i^{f,t}$ and allocate the resource according to $X_i^{w,t}$ \;
		Observe local reward $\psi_i^t$, next state $O_i^{t+1}$ and receive reward feedback from other nodes $\psi_{j\neq i}^t$\;
		Save transition $\langle O_i^t,X_i^t,\psi^t,O_i^{t+1}\rangle$ in $\mathcal{M}_i$\;
		Sample a random mini-batch from $\mathcal{M}_i$ $\mathcal{N}_i=\langle \hat{O}_i^n,X_i^n,\psi^n,\hat{O}_i^{n+1}\rangle$\;
		Set $y_i^n=\psi^n+\gamma\max_{X'\in \mathcal{X}_i}Q_i(O_i^{n+1},X';\theta_{i-}^t)$\;
		Perform a gradient step in (19) with respect to the parameter $\theta_i^t$\;
		Every $\mathcal{C}$ time step, reset the target network parameters $\theta_{i-}^{t+1}=\theta_i^t$\;
		Every $\mathcal{R}^\epsilon$ time step, update $\epsilon_{start}=\delta^\epsilon\cdot\epsilon_{start}$ and $\epsilon_{decay}=-\log{\frac{\epsilon_{min}}{\mathcal{R}^\epsilon}}-\frac{\epsilon_{start}}{\mathcal{R}^\epsilon}$
		
		\textit{t}$\leftarrow$\textit{t}$+1$
	}

	\caption{The deep recurrent Q-learning algorithm for approximating the optimal state-action value functions of a fog node $i\in I$ with experience memory\label{DQR}}
\end{algorithm}

\section{Performance evaluation}\label{sectionfive}
In this section, we quantify the performance gain from the proposed DRQN-based learning algorithm for heterogeneous task offloading and resource allocation problems in multi-fog networks using numerical experiments based on Python-TensorFlow simulator. We used three different environments which are equipped with Inter(R) Core i7-7500 U CPU @ 2.7GHz 64-bit OS, Intel(R) Xeon(R) CPU E3-1225 v6 @ 3.3GHz 64-bit OS, and AMD Ryzen Threadripper 1920X 12-Core Processor.

\subsection{Simulation settings}
For our simulations, we consider a fog layer consisting of five fog nodes that are randomly distributed within a network area of $100\times 100$ $m^{2}$. In addition, a total of three different slices are created on top of each fog node. Slice characteristics are customized by the two-level of heterogeneity, namely the resource demands types and delay constraints, which are summarized in Table \ref{table1}. As an example of slice characteristics in Table \ref{table1}, slice $k$ can be dedicated to Standard resource type services to meet the delay-critical constraint. To obtain realistic values for the processing capacities of fog nodes, we use the CPU processing densities and memory sizes from \cite{b18-1} which used real applications data including a YouTube video data set in \cite{b26-1}. For slice $k$ at fog node $i$, the task arrivals follow a Bernoulli distribution with parameter $\lambda_{i,k}$ (in task/slot) and the packet size is 5$\cdot10^6$ bits. Additionally, the buffer size in each slice is 10, which means that a maximum of 10 slice tasks can stay in the buffer concurrently until processing terminates. The path loss constant and exponent are set to $10^{-3}$ and 4, respectively. The bandwidth for each fog node is 1MHz. The transmission power of the fog node is 20dBm, while the noise power density is -174dBm/Hz \cite{b11}. 
In regard to resource capacity distribution at fog nodes, the CPU speed of a fog node is randomly sampled from [5GHz, 6GHz, 7GHz, 8GHz, 9GHz, 10GHz], where the memory size of a fog node is randomly sampled from [2.4GB, 4GB, 8GB]. The allocation unit of CPU and memory resources are 1GHz and 400MB, respectively. 

\begin{table}[t]
\caption{Two-level of heterogeneity values to service tasks in simulation}
	\label{table1}
	\centering
\begin{tabular}{|c|c|c|c|}
\hline
\multirow{2}{*}{\begin{tabular}[c]{@{}c@{}}\textbf{Resource} \\ \textbf{Types}\end{tabular}} & \multicolumn{3}{c|}{\textbf{Delay Constraints}}                                                                                                                     \\ \cline{2-4} 
                                & Critical                                          & Sensitive                                         & Tolerant                                          \\ \hline
Standard                        & \begin{tabular}[c]{@{}c@{}}$D_k^{max}=10$ms\\ $L^c_k=400$ cycles/bit\\ $L^m_k=400$ Mbytes\end{tabular} & \begin{tabular}[c]{@{}c@{}}50ms\\ 400 cycles/bit\\ 400 Mbytes\end{tabular} & \begin{tabular}[c]{@{}c@{}}100ms\\ 400 cycles/bit\\ 400 Mbytes\end{tabular} \\ \hline
\begin{tabular}[c]{@{}c@{}}CPU \\ intensive\end{tabular}                   & \begin{tabular}[c]{@{}c@{}}$D_k^{max}=10$ms\\ $L^c_k=600$ cycles/bit\\ $L^m_k=400$ Mbytes\end{tabular} & \begin{tabular}[c]{@{}c@{}}50ms\\ 600 cycles/bit\\ 400 Mbytes\end{tabular} & \begin{tabular}[c]{@{}c@{}}100ms\\ 600 cycles/bit\\ 400 Mbytes\end{tabular} \\ \hline
\begin{tabular}[c]{@{}c@{}}Memory \\ intensive\end{tabular}                & \begin{tabular}[c]{@{}c@{}}$D_k^{max}=10$ms\\ $L^c_k=200$ cycles/bit\\ $L^m_k=1200$ Mbytes\end{tabular} & \begin{tabular}[c]{@{}c@{}}50ms\\ 200 cycles/bit\\ 1200 Mbytes\end{tabular} & \begin{tabular}[c]{@{}c@{}}100ms\\ 200 cycles/bit\\ 1200 Mbytes\end{tabular} \\ \hline
\end{tabular}
\end{table}

To evaluate the performance of different neural network settings, three neural networks are considered to estimate the Q-value in our simulation. For all of them, the output layer is a fully connected layer of $|X_i(t)|$ units, where $|X_i(t)|$ represents the dimension of the action set. Additionally,  the activation function of the output layer is a linear activation function, which corresponds to the predicted Q-value of all possible actions. These neural networks differ from each other on the input layer and the hidden layers as detailed below.
\begin{enumerate}
  \item \textit{DRQN}: for the design of the deep recurrent Q-network, the input to the network consists of $Seq\times|O_i(t)|$, where $|O_i(t)|$ is the dimension of the state set and $Seq$ is the sequence size for the 1D-convolutional network. The first hidden layer convolves 32 filters with a kernel size of 3 and applies a Rectified Linear Unit (ReLU). The second hidden layer convolves 64 filters with a kernel size of 3, again followed by a ReLU. This is followed by a recurrent layer in which we use GRU. The number of units in the GRU cell is 128 and the sequence length is 10. The final GRU state is followed by a fully connected layer with ReLU, which has 64 units.
  \item \textit{DCQN}: the deep convolutional Q-network is almost identical to the deep recurrent Q-network except for a recurrent hidden layer. The resulting activations from the second convolutional hidden layer are followed by two fully-connected layers with ReLU, the first one has 128 units and the second has 64 units.
  \item \textit{DQN}: we use four fully-connected hidden layers consisting of 64, 128, 128, and 64 units with ReLU.
\end{enumerate}
In all the experiments, we use the Adam optimizer with a learning rate of 0.001 and learning starts after $10^4$ iterations. A discount factor $\gamma$ of 0.98 is used in the Q-learning update. The replay memory size of $\mathcal{D}_i$ is $10^4$. The target network parameters $\mathcal{C}$ is updated every $10^3$ time slots. We use a mini-batch size of 32 transition samples per update. The $\epsilon$-renewal factor $\delta^{\epsilon}$, $\epsilon$-renewal rate $\mathcal{R}^{\epsilon}$, $\epsilon$-start $\epsilon_{start}$ and minimum-$\epsilon$ $\epsilon_{min}$ are set to 0.9, 5000, 1 and 0.01, respectively. 

For performance comparisons, the existing methods are simulated as baseline schemes. Given the large state and action spaces in the problem considered, we compare methods that are practicable using limited computational resources. Specifically, one baseline offloading method is used as follows:
\begin{itemize}
    \item Threshold Offloading with Nearest Node selection: the node offloads its tasks only if the buffer is above a certain threshold and we set the threshold to 0.8 which implies that the task is to be offloaded if the buffer is more than 80$\%$ full. Also, the node selects the most adjacent neighboring node aiming to minimize communication delays and energy consumption which is an offloading algorithm widely used in IoT and device-to-device communications.
\end{itemize}
On the other hand, two conventional resource allocation algorithms are simulated as baseline schemes, namely:
\begin{itemize}
    \item Round Robin (RR): this algorithm allows every slice that has tasks in the queue to take turns in processing on a shared resource in a periodically repeated order.
    \item Priority Queuing (PQ): this algorithm handles the scheduling of the tasks following a priority-based model. Tasks are scheduled to be processed from the head of a given queue only if all queues of higher priority are empty, which is determined by a delay constraint. 
\end{itemize}
\begin{table}[t]
\caption{Simulation cases according to the three slices' characteristics}
	\label{table2}
	\centering
\small
\begin{tabular}{|c|c|c|c|}
\hline
\multicolumn{1}{|l|}{\textbf{Case}}          & \textbf{fog slice-1}                                              & \textbf{fog slice-2}                                                   & \textbf{fog slice-3}                                                      \\ \hline
\multirow{2}{*}{\textbf{case-1}} & \footnotesize\begin{tabular}[c]{@{}c@{}}Standard\\ Delay-Critical\end{tabular} & \footnotesize\begin{tabular}[c]{@{}c@{}}CPU-intensive\\ Delay-Critical\end{tabular} & \footnotesize\begin{tabular}[c]{@{}c@{}}Memory-intensive\\ Delay-Critical\end{tabular} \\ \cline{2-4} 
                                 & \multicolumn{3}{c|}{\small\begin{tabular}[c]{@{}c@{}}All slices have the same delay constraint,\\ but different resource type tasks\end{tabular}}                                                                            \\ \hline
\multirow{2}{*}{\textbf{case-2}} & \footnotesize\begin{tabular}[c]{@{}c@{}}Standard\\ Delay-Critical\end{tabular} & \footnotesize\begin{tabular}[c]{@{}c@{}}Standard\\ Delay-Sensitive\end{tabular}     & \footnotesize\begin{tabular}[c]{@{}c@{}}Standard\\ Delay-Tolerant\end{tabular}         \\ \cline{2-4} 
                                 & \multicolumn{3}{c|}{\small\begin{tabular}[c]{@{}c@{}}All slices have the same resource type tasks,\\ but different delay priorities\end{tabular}}                                                                            \\ \hline
\multirow{2}{*}{\textbf{case-3}} & \footnotesize\begin{tabular}[c]{@{}c@{}}Standard\\ Delay-Critical\end{tabular} & \footnotesize\begin{tabular}[c]{@{}c@{}}CPU-intensive\\ Delay-Critical\end{tabular} & \footnotesize\begin{tabular}[c]{@{}c@{}}Standard\\ Delay-Sensitive\end{tabular}        \\ \cline{2-4} 
                                 & \multicolumn{3}{c|}{\small\begin{tabular}[c]{@{}c@{}}Some slices have the same resource type tasks,\\ but different delay priorities and some vise versa\end{tabular}}                                                       \\ \hline
\end{tabular}
\end{table}
For different evaluation scenarios, we specifically assign three different cases in terms of slices' characteristics to analyze how each slice's different resource demands and delay priorities are interrelated to each other. Thus, the three simulation cases according to the three slices' characteristics are summarized in Table \ref{table2}. It is noted that this evaluation can simply be expanded by configuring Table \ref{table1} and Table \ref{table2} to suit the needs of the fog network.

\subsection{Performance analysis}
In this subsection, we evaluate the performance of the proposed algorithm by comparing the simulation results under different system parameters.
\subsubsection{Complexity analysis}
In this section, the memory complexity and processing time of the proposed algorithm are investigated. The proposed DRQN-based learning algorithm described in Section \ref{sectionfour}.C requires storing a replay buffer which consists of the state $O^t_i$, action $X^t_i$, reward $\psi^t$, next state $O^{t+1}_i$, and valid action spaces of the next state $\Xi(O^{t+1}_i)$ for the target Q-network. In single transition experience, the state, action, and reward require storing a $\mathbf{len}~A_i+\mathbf{len}~B_i+\mathbf{len}~B^e_i+\mathbf{len}~R_i=3\times K\times2$ size array of decimal values, single integer numbers, and single float numbers, respectively. Moreover, the valid action space of the next state is an $|\mathcal{X}_i|$ size array of binary values, where $\Xi[X']=0$ if $X'\in\mathcal{X}_i$ is invalid in state $O^{t+1}_i$, otherwise  $\Xi[X']= 1$. Finally, using the parameters in Section \ref{sectionfour}.A, the proposed algorithm requires much less memory than the conventional Q-learning algorithm, i.e., approximately 2.8GB compared to 7.24TB. It is worth mentioning that we leverage a Python-based simulator where the array as a whole is an object that stores the float data in consecutive regions of the memory, and thus the memory size of an individual float is not explicit. Therefore, the memory usage compared corresponds to the data contained in the object.

Furthermore, multi-node learning alleviates the overhead of the network infrastructure as well as improves the system response time, compared to the centralized architecture. In our simulation, assuming each node is equipped with a single CPU, the processing time per iteration is around 0.04s. Moreover, the proposed neural network model can be trained in parallel on multiple CPUs or GPUs to improve the training time and memory usage.

\subsubsection{Convergence performance}
\begin{figure*}
  \centering\includegraphics[width=1\linewidth]{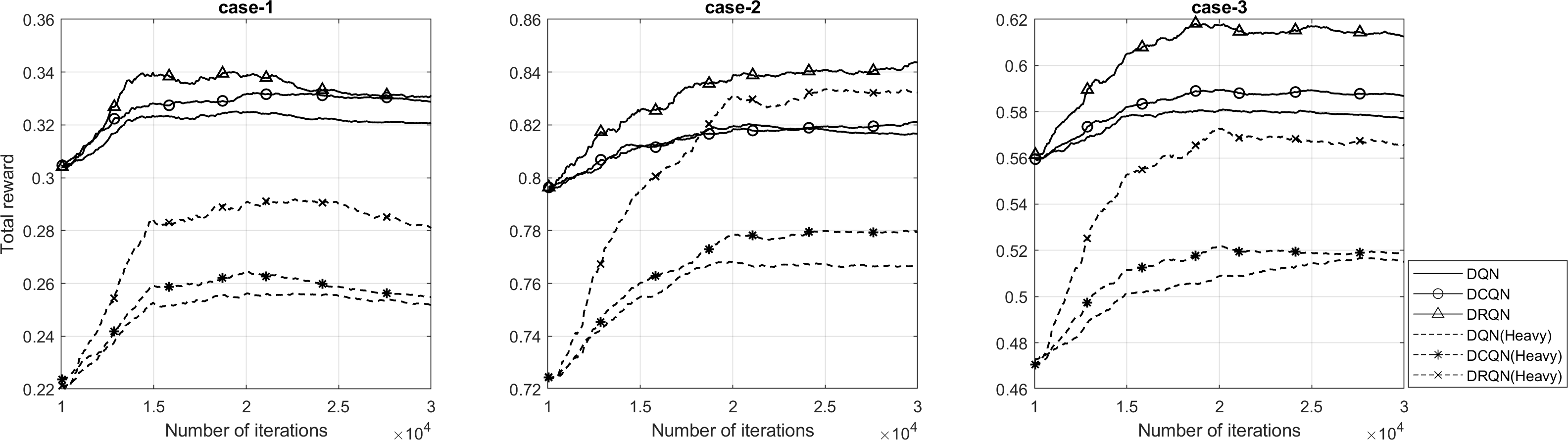}\par
  \caption{The convergence property of the proposed algorithm using different neural networks.}
  \label{figure3}
\end{figure*}

In this experiment, we evaluate the convergence property of different neural networks with the above parameter settings to confirm whether the proposed deep reinforcement learning-based algorithm can achieve stable performance. To quantify the impact of task traffic status on the convergence performance, we implement two different average task arrival rates, which are categorized into normal($\bar{\lambda}$=0.6) and heavy($\bar{\lambda}$=0.8), where $\bar{\lambda}$ is the average task arrival rate per slice at the fog node. Since the uniform random policy runs for $10^4$ iterations at all nodes before learning starts, the total average reward value is not enhanced during this time and thus we show the average total reward of fog nodes after they start learning their networks. Once each node starts learning its own state-action value functions with a preassigned neural network, the total average rewards are increasing and asymptotically converge after around $1.5\times10^4$ iterations as shown in Fig. \ref{figure3}. In regard to the average task arrival rate, when the nodes receive a smaller number of tasks per time slot, the average total reward value is larger over all simulation cases. The main reason behind this is that the number of successfully processed tasks with limited resources of fog nodes is higher when a fewer number of tasks are waiting in the buffer and also that the buffers are less likely to be overflowed. Given the findings from this experiment, the proposed algorithm using DRQN can achieve greater total reward compared to DQN and DCQN. This result implies that DRQN controllers can handle partial observability by retaining some memory of previous observations to help the nodes achieve better decision making.

\subsubsection{Performance under different characteristics of fog slices}
\begin{figure}[t]
	\centering
	\includegraphics[width=0.8\linewidth]{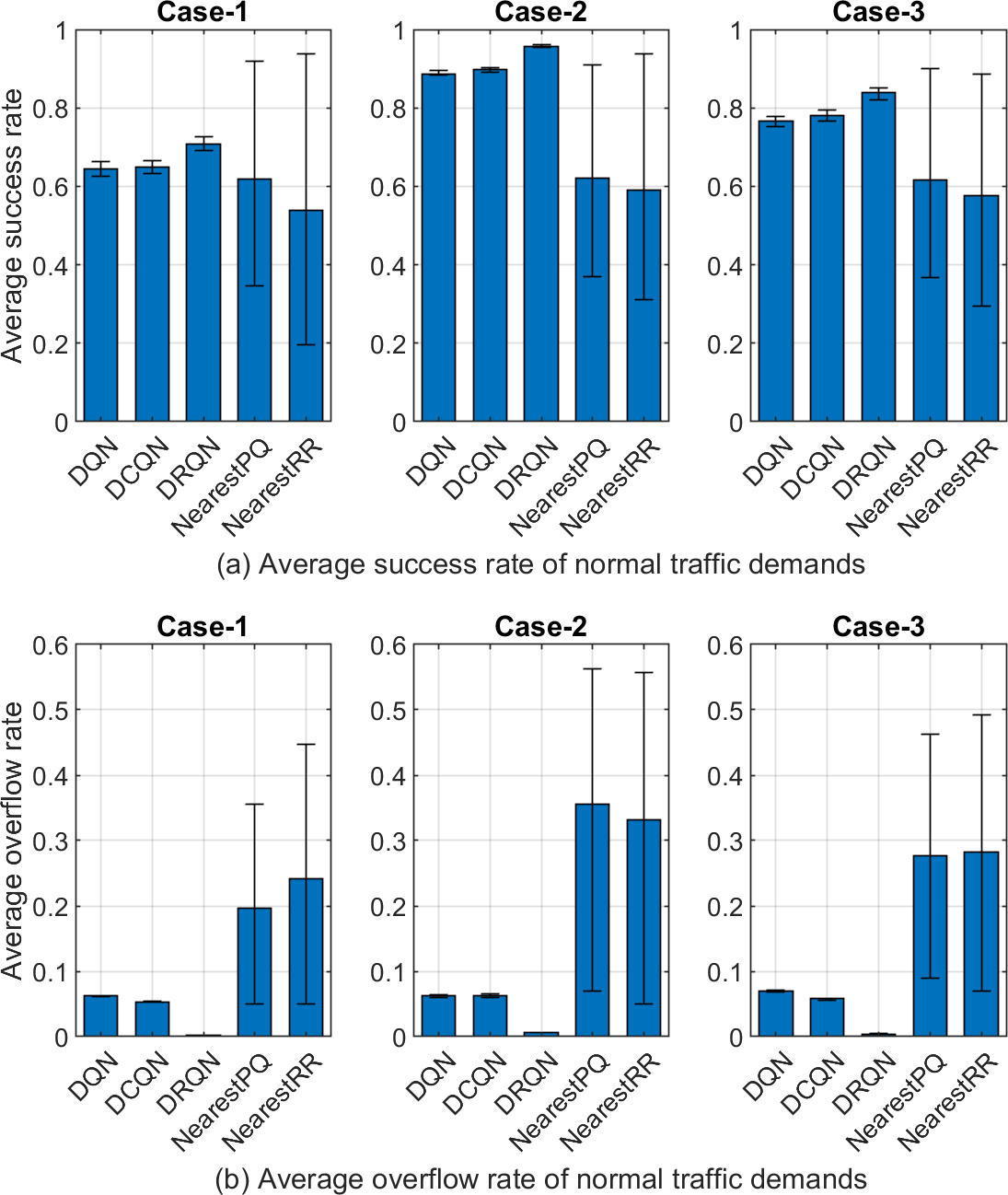}
	\caption{Performance of (a) average success rate and (b) average overflow rate of normal traffic under different cases.}
	\label{normalall}
\end{figure}

\begin{figure}[t]
	\centering
	\includegraphics[width=0.8\linewidth]{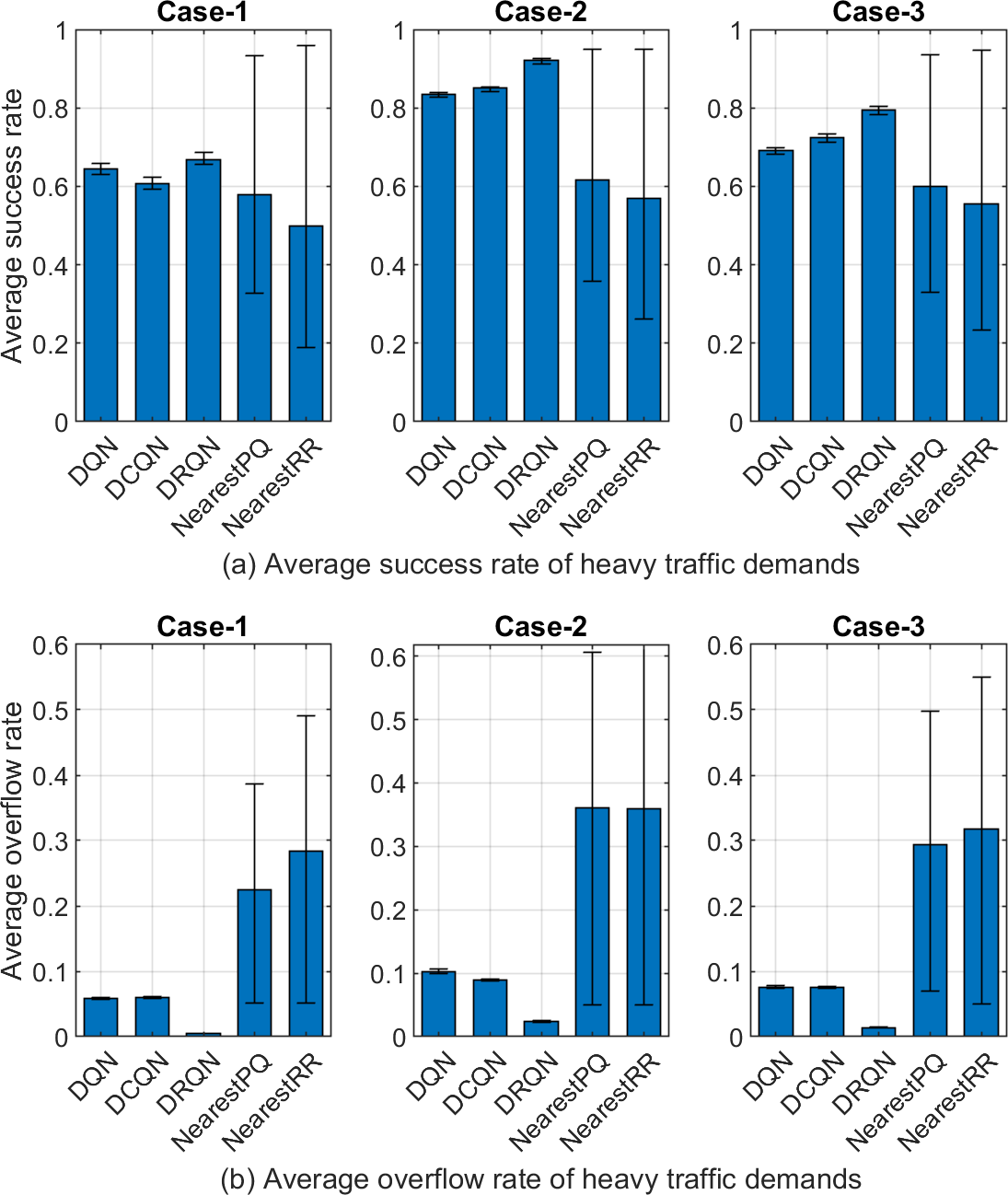}
	\caption{Performance of (a) average success rate and (b) average overflow rate of heavy traffic under different cases.}
	\label{heavyall}
\end{figure}

This experiment mainly aims to demonstrate the performance in terms of the average processing success rate, the average overflow rate, and the average task latency under different characteristics of fog slices as shown in Table \ref{table2}. Like the previous experiment, two kinds of traffic demands, normal and heavy, are used for evaluation. Fig. \ref{normalall} and Fig. \ref{heavyall} illustrates the average success rate of tasks and the average overflow rate per fog node. From Fig. \ref{normalall} (a) and Fig. \ref{heavyall} (a), it can be observed that the proposed scheme using DRQN achieves the highest average success rate. Moreover, as the traffic requested increases, a larger number of tasks fail to attain their delay performance requirements due to the lack of resources. Comparing the three cases, when the nodes are requested the same traffic rate but demanding high computation and memory resources (Case-1), fog nodes are more likely to experience a lower success rate. This is because the processing time takes longer with limited resources, which also leads to failure of the delay requirements.

In Fig. \ref{normalall} (b) and Fig. \ref{heavyall} (b), it is shown that, when implementing baseline methods, extremely large amounts of tasks are dropped due to overloading buffers. This is because fog nodes always select the most proximate fog node to offload their tasks, where the same fog node can be selected by several neighbors and thus its buffers will fill up quickly with tasks from multiple neighboring nodes. Furthermore, the resource allocation methods also affect the overflow performance. As we mentioned, the slices with large resource demands take more processing time than the slices with small resource demands. Thus, when implementing RR and PQ methods, there is unfairness in the allocation of slices with small resource demands, which induces the high number of task drops from overloaded buffers. On the other hand, even though slices of Case-2 constitute the tasks with the same resource demands, their overflow rate is higher than that of Case-1 and Case-3. The difference in Case-2 is that the fog slice-3 of nodes are dedicated to delay-tolerant tasks where tasks of this slice can stay in the buffer until they exceed their large delay limit. Hence, the average processing success rate from Case-2 is higher due to relatively adequate delay limits, while the average overflow performance is worse. However, the proposed algorithm can offload their tasks to different neighboring nodes depending on their buffer and resource status and avoid unfairness among slices with different priorities in the resource allocation process, leading to an increased average success rate. 
Moreover, Fig. \ref{normalall} and Fig. \ref{heavyall} show that the variances of the success and overflow rates indicated by the error bars vary from one algorithm to the other. The error bars in these figures represent the largest value as the upper limit and the smallest value as the lower limit among all the nodes. For example in Fig. \ref{normalall} (a), the success rate of Case-2 using DRQN has a mean of 95.6$\%$ and varies between 95.3$\%$ to 96.1$\%$, while the success rate of Case-2 using nearest node selection with PQ resource allocation has a mean of 62.1$\%$ and varies between 36.8$\%$ to 90.8$\%$. We can clearly see that the variability of the task success rate between fog nodes is greater for baseline methods than for the proposed algorithms, where the same trend is shown in the overflow rate. This result indicates that the proposed algorithm discourages selfish behavior in nodes and achieves a win-win cooperation between fog nodes by making rational offloading and resource allocation decisions.

\begin{figure*}
  \centering\includegraphics[width=0.65\linewidth]{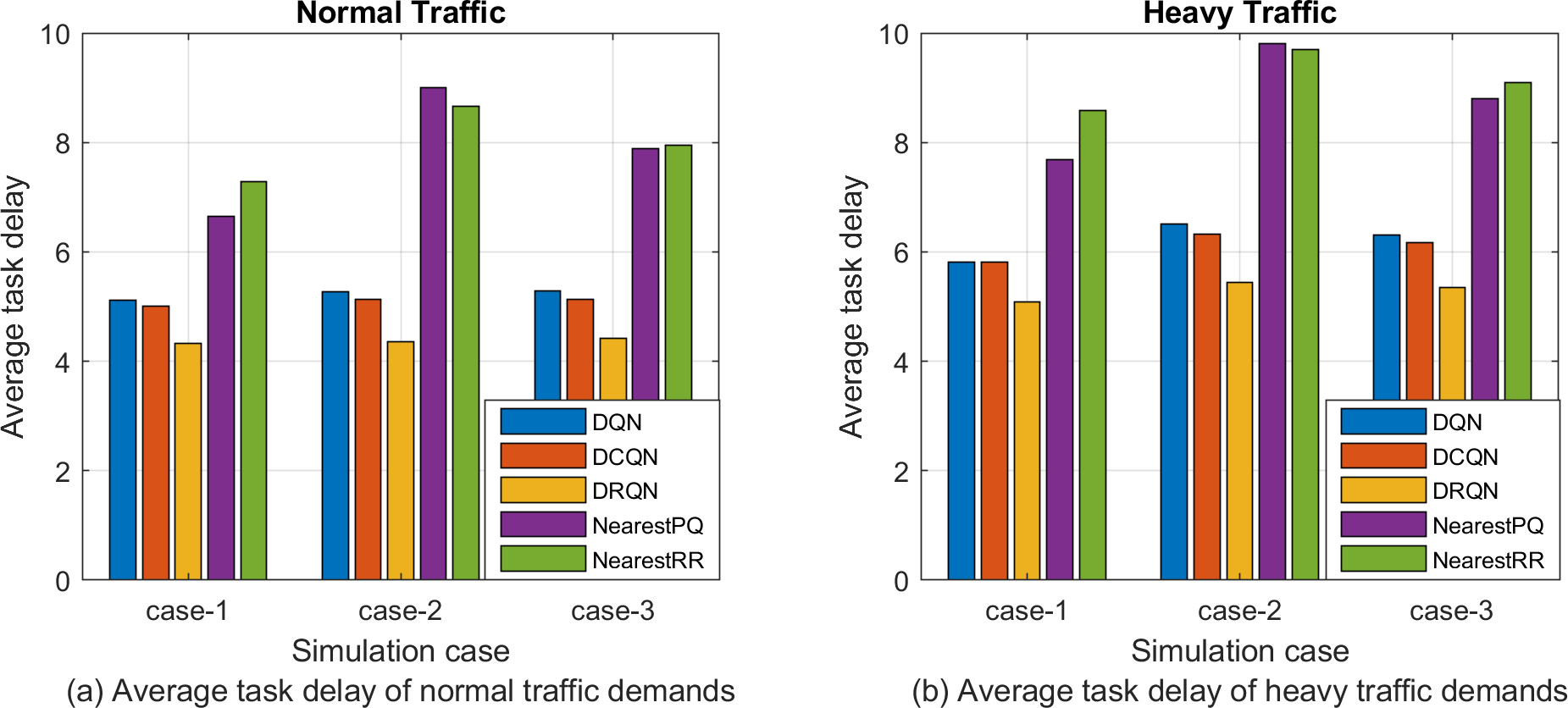}
  \caption{Task delay performance of (a) normal traffic and (b) heavy traffic under different cases.}
  \label{figure5}
\end{figure*}

Fig. \ref{figure5} illustrates the average task delay under different cases. In contrast to the baseline methods, the proposed algorithm decreases the average task delay by selecting a neighboring fog node that minimizes transmission delay as well as waiting time in the buffers and allowing distinct resource allocation with respect to characteristics of each slice. 

\subsubsection{Performance under different task arrival rates}
\begin{figure*}
  \centering\includegraphics[width=0.9\linewidth]{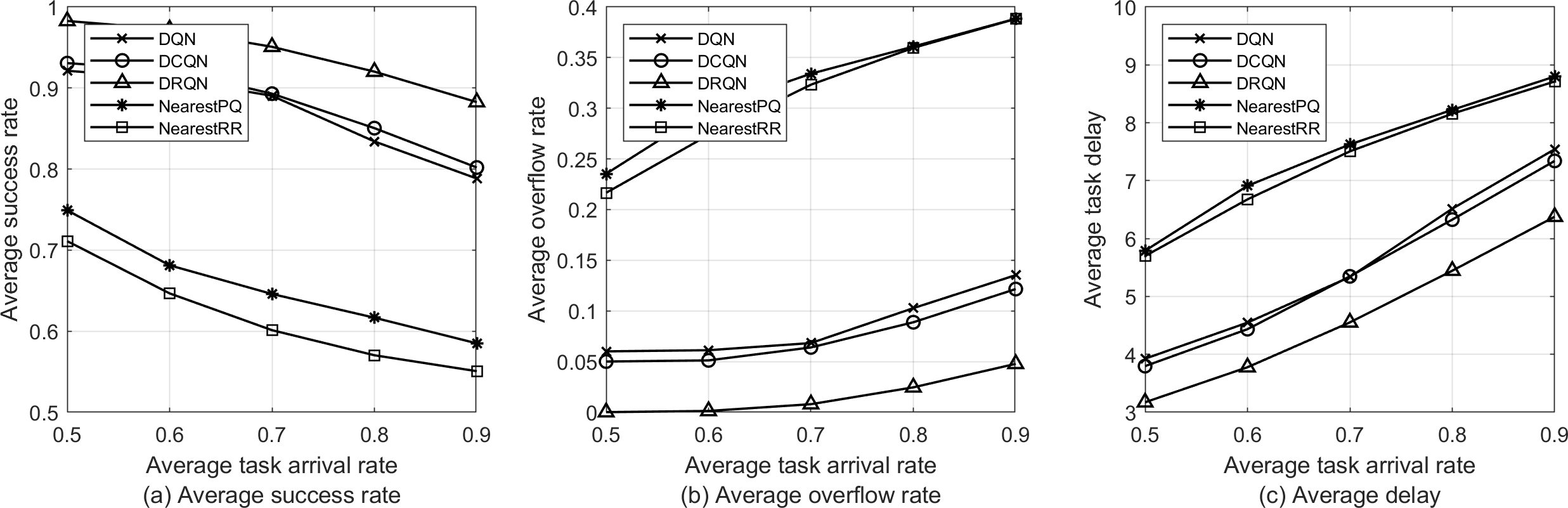}
  \caption{Performance of (a) average success rate, (b) average overflow rate, and (c) average task delay of Case-2 under different task arrival rates.}
  \label{figure6}
\end{figure*}

In this experiment, we consider how the task arrival rates impact the average performance in terms of the average processing success rate, the average overflow rate, and the average task latency per slice. In this simulation, we set all fog nodes' slice characteristics to Case-2. In Fig. \ref{figure6} (a), as the task arrival rate increases, more tasks fail to be completed within their delay limits. Similar observations can be found from Fig. \ref{figure6} (b) where more tasks are dropped when the task arrival rate increases from 0.5 to 0.9. The reason behind this is that, as the task arrival rate increases, the waiting time becomes longer due to the larger number of tasks waiting in the buffer, which means that the tasks are more likely to fail their delay requirements or be dropped if they arrive when the buffer is full. Meanwhile, over the variation of the task arrival rate, the maximum success rate and minimum overflow rate is achieved from the proposed algorithm. Moreover, as shown in Fig. \ref{figure6} (c), the DRQN-based algorithm has the lowest average delay in Case-2. 
These empirical results show that temporal integration of observations from a recurrent network allows the nodes to coordinate in their choices without knowing the explicit state and action sets of the others which makes the proposed DRQN-based algorithm relatively robust to the dynamics of the partially observable environment. These results also demonstrate that intelligently distributing resources to slices requiring different delay constraints makes a huge impact on the overall system performance.

\section{Conclusion}\label{sectionsix}
In this paper, we devised a joint heterogeneous task offloading and resource allocation algorithm whose goal is to maximize the processing tasks completed within their delay constraints while minimizing the task drops from buffer overflows. The SDN-based fog network we consider has multiple fog nodes that are coordinating to achieve the best overall network performance without knowing the explicit status of other fog nodes. In the presence of uncertainties stemming from task demands and resource status, we formulate the problem as a partially observable stochastic game and apply cooperative multi-agent deep reinforcement learning with a global reward that aims to maximize the common goal of nodes and stabilize the convergence property. Further, we implement a recurrent neural network to tackle the partial-observability by maintaining internal states and aggregating temporal observations. The simulation results show that the proposed DRQN-based algorithm can achieve a higher average success rate and lower average overflow than DQN and DCQN as well as non-deep learning based baseline methods. In the future, we will extend the multi-agent learning to scenarios for agents in large-scale fog networks with differing reward functions.

\begin{IEEEbiography}
[{\includegraphics[width=1in,height=1.25in,clip,keepaspectratio]{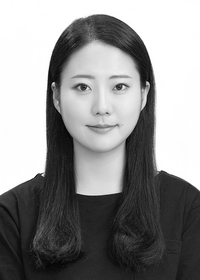}}]{Jungyeon Baek}
(Member, IEEE) received the bachelor and the M.S. degree in Electronics and Radio Engineering from KyungHee University, Yongin, South Korea, in the year of 2015 and 2017, respectively. She is currently working toward the Ph.D. degree in Electrical Engineering at École de Technologie Supérieure (ETS), University of Québec, Montr\'eal, Canada. Her research interests include fog/edge computing, resource management, QoS provisioning in computation and communication networks, software-defined networking, reinforcement learning, and deep learning. Her research findings are published in many prestigious venues such as IEEE International Symposium on Personal Indoor and Mobile Radio Communications (PIMRC) and IEEE Wireless Communications and Networking Conference (WCNC). 
\end{IEEEbiography}

\begin{IEEEbiography}
[{\includegraphics[width=1in,height=1.25in,clip,keepaspectratio]{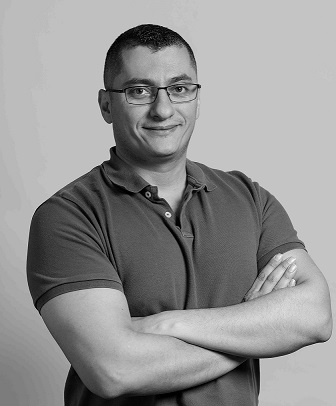}}]{Georges Kaddoum}
(Member, IEEE) received the Bachelor’s degree in electrical engineering from the \'Ecole Nationale Supérieure de Techniques Avancées (ENSTA Bretagne), Brest, France, and the M.S. degree in telecommunications and signal processing(circuits, systems, and signal processing) from the Universit\'e de Bretagne Occidentale and Telecom Bretagne (ENSTB), Brest, in 2005 and the Ph.D. degree (with honors) in signal processing and telecommunications from the National Institute of Applied Sciences (INSA), University of Toulouse, Toulouse, France, in 2009. He is currently an Associate Professor and Tier 2 Canada Research Chair with the \'Ecole de Technologie Sup\'erieure (\'ETS), Universit\'e du Qu\'ebec, Montr\'eal, Canada. In 2014, he was awarded the \'ETS Research Chair in physical-layer security for wireless networks.  Since 2010, he has been a Scientific Consultant in the field of space and wireless telecommunications for several US and Canadian companies. He has published over 200+ journal and conference papers and has two pending patents. His recent research activities cover mobile communication systems, modulations, security, and space communications and navigation. Dr. Kaddoum received the Best Papers Awards at the 2014 IEEE International Conference on Wireless and Mobile Computing, Networking, Communications (WIMOB), with three coauthors, and at the 2017 IEEE International Symposium on Personal Indoor and Mobile Radio Communications (PIMRC), with four coauthors. Moreover, he received IEEE Transactions on Communications Exemplary Reviewer Award for the year 2015, 2017, 2019. In addition, he received the research excellence award of the Universit\'e du Qu\'ebec in the year 2018. In the year 2019, he received the research excellence award from the \'ETS in recognition of his outstanding research outcomes. Prof. Kaddoum is currently serving as an Associate Editor for IEEE Transactions on Information Forensics and Security, and IEEE Communications Letters.
\end{IEEEbiography}


\begin{thebibliography}{00}
\bibitem{b0} A. Zaidi, Y. Hussain, M. Hogan, and C. Kuhlins, "Cellular IoT evolution for industry digitalization," Ericsson, White Paper, 2019. [Online]. Available: https://www.ericsson.com/en/reports-and-papers/white-papers/cellular-iot-evolution-for-industry-digitalization 
\bibitem{b1} M. Peng, S. Yan, K. Zhang, and C. Wang, “Fog-computing-based radio access networks: Issues and challenges,” \textit{IEEE Network}, vol. 30, no. 4, Jul. 2016, pp. 46–53.
\bibitem{b2} M. Chiang and T. Zhang, “Fog and IoT: An overview of research opportunities,” \textit{IEEE Internet of Things J.}, vol. 3, Dec. 2016, pp. 854–864. 
\bibitem{b3} Y. J. Ku, D. Y. Lin, C. F. Lee, P. J. Hsieh, H. Y. Wei, C. T. Chou, and A. C. Pang, “5G radio access network design with fog paradigm: Confluence of communications and computing,” \textit{IEEE Commun. Mag.}, Apr. 2017.
\bibitem{b4} C. Mouradian, D. Naboulsi, S. Yangui, R. H. Glitho, M. J. Morrow, and P. A. Polakos, “A comprehensive survey on fog computing: Stateof-the-art and research challenges,” \textit{IEEE Commun. Surveys Tut.}, vol. 20, no. 1, Nov. 2017, 416-464.
\bibitem{b5} A. V. Dastjerdi, and R. Buyya, \textit{Internet of Things principles and paradigms}, Elsevier, 2016. [E-book] 
\bibitem{b6} A. A. Fuqaha, M. Guizani, M. Mohammadi, M. Aledhari, and M. Ayyash, “Internet of things: A survey on enabling technologies, protocols, and applications,” \textit{IEEE Commun. Surveys Tuts.}, vol. 17, no. 4, 2015.
\bibitem{b11} G. Lee, W. Saad, and M.Bennis, “An online optimization framework for distributed fog network formation with minimal latency,” arXiv preprint arXiv:1710.05239, 14 Oct 2017.
\bibitem{b11-1} X. Xu, S. Fu, Q. Cai, W. Tian, W. Liu, W. Dou, X. Sun, A. Liu, "Dynamic resource allocation for load balancing in fog environment", \textit{Wireless Communications and Mobile Computing}, pp. 1-15, 2018.
\bibitem{b11-2} W. Zhang, Z. Zhang, and H. Chao, “Cooperative fog computing for dealing with big data in the internet of vehicles: Architecture and hierarchical resource management,” \textit{IEEE Commun. Mag.}, vol. 55, no. 12, pp. 60–67, 2017.
\bibitem{b9-2} M. Aazam, S. Zeadally, K. A. Harras, "Offloading in fog computing for IoT: Review, enabling technologies, and research opportunities," \textit{Future Generation Comp. syst.}, vol.
87, Oct. 2018, pp. 278-289.
\bibitem{b9-3} P. Mach and Z. Becvar, "Mobile edge computing: A survey on architecture and computation offloading," in \textit{IEEE Commun. Surv. Tut.}, vol. 19, no. 3, 2017.
\bibitem{b9-4} L. Yu, L. Chen, Z. Cai, H. Shen, Y. Liang, and Y. Pan , “Stochastic load balancing for virtual resource management in datacenters,” in \textit{IEEE Trans. Cloud Comput.}, Nov. 2014.
\bibitem{b9-5} J. Y. Baek, G. Kaddoum, S. Garg, K. Kaur, and V. Gravel, “Managing fog networks using reinforcement learning based load balancing algorithm,” in \textit{2019 IEEE Wireless Communications and Networking Conference (WCNC)}, 15- 18 Apr. 2019.
\bibitem{b12} A. Yousefpour, G. Ishigaki, R. Gour, and J.P. Jue, "On reducing IoT service delay via fog offloading," \textit{IEEE Internet of Things J.}, vol. 5, no. 2, Apr. 2018, pp. 998-1010.
\bibitem{b13} H. Zhang, Y. Xiao, S. Bu, D. Niyato, F.R. Yu, and Z. Han, "Computing resource allocation in three-tier IoT fog networks: A joint optimization approach combining stackelberg game and matching," \textit{IEEE Internet of Things J.}, vol. 4, no. 5, Oct. 2017, pp. 1204-1215.
\bibitem{b14} C. Wang, C. Liang, F.R. Yu, Q. Chen, and L. Tang, "Computation offloading and resource allocation in wireless cellular networks with mobile edge computing," in \textit{IEEE Trans. Wireless Commun.}, vol. 16, no. 8, Aug. 2017, pp. 4924-4938.
\bibitem{b15} H. A. Alameddine, S. Sharafeddine, S. Sebbah, S. Ayoubi, and C. Assi, "Dynamic task offloading and scheduling for low-latency IoT services in multi-access edge computing," in \textit{IEEE J. Sel. Areas Commun.}, vol. 37, no. 3, Mar. 2019, pp. 668-682.
\bibitem{b16} L. Li, Q. Guan, L. Jin, and M. Guo, "Resource allocation and task offloading for heterogeneous real-time tasks with uncertain duration time in a fog queueing system," in \textit{IEEE Access}, vol. 7, 2019, pp. 9912–9925.
\bibitem{convex} Luo ZQ and Yu W, “An introduction to convex optimization for communications and signal processing,” \textit{IEEE J. Select. Areas Commun.}, vol.24, no.8, Aug. 2008, pp.1426–1438.
\bibitem{m1} S. K. Mishra, D. Puthal, J. J. P. C. Rodrigues, B. Sahoo, and E. Dutkiewicz, "Sustainable service allocation using metaheuristic technique in fog server for industrial applications," \textit{IEEE Trans. Ind. Informat.}, vol. 14, no. 10, Oct. 2018.
\bibitem{m2} C. W. Tsai and J. J. Rodrigues, “Metaheuristic scheduling for cloud: A survey,” \textit{IEEE Syst. J.}, vol. 8, no. 1, pp. 279–291, Mar. 2014.
\bibitem{m3} N. Bergmann, Y. Y. Chung, and X. Yang, “Using swarm intelligence to optimize the energy consumption for distributed systems,” \textit{Mod. Appl. Sci.}, vol. 7, no. 6, pp. 59–66, 2013.
\bibitem{m4} D. Zhang, F. Haider, M. S. Hilaire, and C. Makaya, "Model and algorithms for the planning of fog computing networks," \textit{IEEE Internet of Things J.}, vol. 6, no. 2, Apr. 2019.
\bibitem{r1} J. Wang, C. Jiang, H. Zhang, Y. Ren, K. C. Chen, and L. Hanzo, "Thirty years of machine learning: The road to pareto-optimal wireless networks." \textit{IEEE Commun. Surveys Tuts.}, early access, Jan. 13, 2020.
\bibitem{r2} C. Zhang, P. Patras, and H. Haddadi, "Deep learning in
mobile and wireless networking: A survey," \textit{IEEE Commun. Surveys Tuts.}, Mar. 2019.
\bibitem{r3} Y. Sun, M. Peng, Y. Zhou, Y. Huang, and S. Mao, “Application
of machine learning in wireless networks: Key techniques and open
issues,” \textit{IEEE Commun. Surveys Tuts.}, vol. 21, no. 4, 2019, pp. 3072-3108.
\bibitem{b20} R. S. Sutton and A. G. Barto, \textit{Introduction to Reinforcement learning} 2nd edition. Cambridge, MA, USA: MIT Press, 1998.
\bibitem{b26} V. Mnih, K. Kavukcuoglu, D. Silver, A. A. Rusu, J. Veness, M. G. Bellemare, A. Graves, M. Riedmiller, A. K. Fidjeland, G. Ostrovski, S. Petersen, C. Beattie, A. Sadik, I. Antonoglou, H. King, D. Kumaran, D. Wierstra, S. Legg, and D. Hassabis, “Human-level control through deep reinforcement learning,” \textit{Nature}, vol. 518, no. 7540, pp. 529–533, Feb. 2015.
\bibitem{r3-1} N. C. Luong, D. T. Hoang, S. Gong, D. Niyato, P. Wang, Y. C. Liang, and D. I. Kim, "Applications of deep reinforcement learning in communications and networking: A survey," \textit{IEEE Commun. Surveys Tuts.}., vol. 21, no. 4. May. 2019, pp. 3133-3174
\bibitem{r3-2} X. Chen, H. Zhang, C. Wu, S.Mao, Y. Ji, and M. Bennis, "Performance optimization in mobile-edge computing via deep reinforcement learning," arXiv preprint arXiv:1804.00514, Mar. 2018.
\bibitem{r3-3} S. Pan, Z. Zhang, Z. Zhang, and D. Zeng, " Dependency-aware computation offloading in mobile edge computing: A reinforcement learning approach," \textit{IEEE Access}, vol. 7, Sep. 2019.
\bibitem{r8} N. V. Huynh, D. T. Hoang, D. N. Nguyen, and E. Dutkiewicz, “Optimal and fast real-time resource slicing with deep dueling neural networks,” \textit{IEEE J. Sel. Areas Commun.}, vol. 37, no. 6, Jun. 2019.
\bibitem{r9} X. Chen, Z. Zhao, C. Wu, M. Bennis, H. Liu, Y. Ji, and H. Zhang ,"Multi-tenant cross-slice resource orchestration: A deep reinforcement learning approach”, \textit{IEEE J. Sel. Areas Commun.}, vol. 37, no. 10, pp. 2377-2392, Aug. 2019.
\bibitem{r10} Y. Sun, M. Peng, and S. Mao, "Deep reinforcement learning based mode selection and resource management for green fog radio access networks," \textit{IEEE Internet of Things J.}, vol. 6, no. 2, pp. 1960-1971, Apr. 2019.
\bibitem{r5} A. Sadeghi, G. Wang, and G. B. Giannakis, "Deep reinforcement learning for adaptive caching in hierarchical content delivery networks," \textit{IEEE Trans. Cogn. Commun. Netw.}, vol. 5, no. 4, Dec. 2019.
\bibitem{r4} Y. Sun, M. Peng, S. Mao, "A game-theoretic approach to cache and radio resource management in fog radio access networks," \textit{IEEE Trans. Veh. Technol.}, vol. 68, Oct. 2019.
\bibitem{r7} Y. He, C. Liang, F. R. Yu, N. Zhao, and H. Yin, “Deep-reinforcement-learning-based optimization for cache-enabled opportunistic interference alignment wireless networks,” \textit{IEEE Trans. Veh. Technol.}, vol. 66, no. 11, Nov. 2017.
\bibitem{r6} K. N. Doan, M. Vaezi, W. Shin, H. V. Poor, H. Shin, and T. Q. S. Quek, "Power allocation in cache-aided NOMA systems: Optimization and deep reinforcement learning approaches," \textit{IEEE Trans. Commun.}, doi: 10.1109/TCOMM.2019.2947418.
\bibitem{b9} R. Jain and S. Paul, “Network virtualization and software defined networking for cloud computing: A survey,” \textit{IEEE Commun. Mag.}, vol. 51, no. 11, Nov. 2013, pp. 24–31.
\bibitem{b6-1} N. McKeown, “Software-defined networking,” \textit{INFOCOM Keynote Talk}, vol. 17, no. 2, 2009, pp. 30–32.
\bibitem{b6-2} K. Mahmood, A. Chilwan, O. Østerbø , and M. Jarschel, “Modelling of OpenFlow-based software-defined networks: The multiple node case,” \textit{IET Netw.}, vol. 4, no. 5, 2015, pp. 278–84.
\bibitem{b8-1} S. Tomovic, K. Yoshigoe, I. Maljevic, and I. Radusinovic., “Software-defined fog network architecture for IoT,” \textit{Wireless Personal Commun.}, Springer, 2016.
\bibitem{b17}  D. Gupta and R. Jahan, “Inter-SDN controller communication: Using border gateway protocol”, \textit{White Paper by Tata Consultancy Services (TCS)}, Jun. 2014.
\bibitem{b17-1} X. Hou, W. Muqing, L. Bo, and L. Yifeng, "Multi-controller deployment algorithm in hierarchical architecture for SDWAN", \textit{IEEE Access}, vol. 7, pp. 65839-65851, 2019.
\bibitem{b18} P. Wang, S.C. Lin, and M. Luo, "A framework for QoS-aware traffic classification using semi-supervised machine learning in SDNs," in \textit{IEEE Int. Conf. on Services Computing (SCC)}, Jun. 2016, CA, USA. pp.760-765.
\bibitem{b18-1} J. Kwak, O. Choi, and S. Chong, "Processor-network speed scaling for energy–delay tradeoff in smartphone application," in \textit{IEEE/ACM Trans. Netw.}, vol. 24, no. 3, Jun. 2016.
\bibitem{b19} M. Verma, M. Bhardawaj, and A. K. Yadav, "An architecture for load balancing techniques for fog computing environment," \textit{Int. J. Computer Sci. and Commun.}, vol. 6, no. 2, Apr./Sep. 2015, pp. 269-274.
\bibitem{b21} L. Busoniu, R. Babuska, B. De Schutter, "A comprehensive survey of multiagent reinforcement learning", in \textit{IEEE Trans. Syst. Man Cybern.}, vol. 38, no. 2, pp. 156-172, 2008.
\bibitem{b22} R. Lowe, Y. Wu, A. Tamar, J. Harb, P. Abbeel, I. Mordatch, "Multi-agent actor-critic for mixed cooperative-competitive environments", \textit{CoRR}, vol. abs/1706.02275, 2017.
\bibitem{b23} M. Hausknecht, P. Stone, "Deep recurrent Q-learning for partially observable MDPs", \textit{Association for the Advancement of Artificial Intelligence Fall Symp. Series}, 2015.
\bibitem{b23-1} C. Claus and C. Boutilier, “The dynamics of reinforcement learning in cooperative multiagent systems,” in \textit{Proc. 15th Nat. Conf. Artif. Intell. 10th Conf. Innov. Appl. Artif. Intell. (AAAI/IAAI-98)}, Madison, WI, Jul. 26–30, pp. 746–752.
\bibitem{b23-2} S. Kapetanakis and D. Kudenko, “Reinforcement learning of coordination in cooperative multi-agent systems,” in \textit{Proc. 18th Nat. Conf. Artif. Intell. 14th Conf. Innov. Appl. Artif. Intell. (AAAI/IAAI-02)}, Menlo Park, CA, Jul. 28–Aug. 1, pp. 326–331.
\bibitem{b23-3} D. Fudenberg and D. M. Kreps, \textit{Lectures on Learning and Equilibrium in Strategic Form Games}, CORE Foundation, Louvain-La-Neuve, Belgium, 1992.
\bibitem{b24} R. E. Bellman, \textit{Dynamic Programming}, Princeton, NJ, Princeton University Press, 1957. 
\bibitem{b25} M. Puterman, \textit{Markov Decision Processes: Discrete Stochastic Dynamic Programming}, New York, NY: John Wiley \& Sons, Inc., 1994.
\bibitem{b26-1} Dataset for Statistics and Social Network of YouTube Videos. [Online]. Available: http://netsg.cs.sfu.ca/youtubedata/
\bibitem{b27} S. Omidshafiei, J. Pazis, C. Amato, J. P. How, and J. Vian, "Deep decentralized multi-task multi-agent reinforcement learning under partial observability", in \textit{Proc. Int. Conf. Mach. Learn. (ICML)}, pp. 2681-2690, 2017.
\bibitem{b28} J. Chung, C. Gulcehre, K. Cho, and Y. Bengio, "Empirical evaluation of gated recurrent neural networks on sequence modeling," \textit{CoRR}. 2014. [Online]. Available: http://arxiv.org/abs/1412.3555
\bibitem{b29} R. Jozefowicz, W. Zaremba, and l. Sutskever, "An empirical exploration of recurrent network architectures," in \textit{Proc. 32nd Int. Conf. on Machine Learning}, vol. 37, Jul. 2015, pp.2342-2350.
\end{thebibliography}
\end{document}